\documentclass[
aip,
mph,
amsmath,amssymb,
superscriptaddress,
floatfix,
preprint,
eqsecnum,
numerical,
]{revtex4-2}
\usepackage{graphicx}
\usepackage{amsmath}
\usepackage{cases}
\usepackage{subcaption}
\usepackage{amsthm,amssymb}
\usepackage{mathrsfs}
\usepackage{epsfig}
\usepackage{epstopdf}
\usepackage{lipsum}
\usepackage{wrapfig}
\usepackage{makecell}
\usepackage{tabularx}
\usepackage{multirow}
\usepackage{soul}
\usepackage{booktabs}
\usepackage{lmodern}
\usepackage{amsmath}
\usepackage[OT2,T1]{fontenc}
\usepackage[bookmarks=false]{hyperref}
\usepackage{color}
\usepackage{float}
\hypersetup{pdfauthor={},pdfborder=0 0 0,colorlinks=true,citecolor=blue,linkcolor=blue,urlcolor=blue,}

\begin{document}
	\title{Matrixed-Spectrum Decomposition Accelerated Linear Boltzmann Transport Equation Solver for Fast Scatter Correction in Multi-Spectral CT}
	\author{Guoxi Zhu}
	\affiliation{Department of Engineering Physics, Tsinghua University, Beijing 100084, China} 
	\affiliation{Key Laboratory of Particle and Radiation Imaging, Ministry of Education, Beijing 100084, China}
	\author{Li Zhang}
	\affiliation{Department of Engineering Physics, Tsinghua University, Beijing 100084, China} 
	\affiliation{Key Laboratory of Particle and Radiation Imaging, Ministry of Education, Beijing 100084, China}
	\author{Zhiqiang Chen}
	\email{czq@mails.tsinghua.edu.cn}
	\affiliation{Department of Engineering Physics, Tsinghua University, Beijing 100084, China} 
	\affiliation{Key Laboratory of Particle and Radiation Imaging, Ministry of Education, Beijing 100084, China}
	\author{Hewei Gao}
	\email{hwgao@tsinghua.edu.cn}
	\affiliation{Department of Engineering Physics, Tsinghua University, Beijing 100084, China} 
	\affiliation{Key Laboratory of Particle and Radiation Imaging, Ministry of Education, Beijing 100084, China}

	\begin{abstract}
		Abstract: X-ray scatter has been a serious concern in computed tomography (CT), leading to image artifacts and distortion of CT values. The linear Boltzmann transport equation (LBTE) is recognized as a fast and accurate approach for scatter estimation. However, for multi-spectral CT, it is cumbersome to compute multiple scattering components for different spectra separately when applying LBTE-based scatter correction. 
		In this work, we propose a Matrixed-Spectrum Decomposition accelerated LBTE solver (MSD-LBTE) that can be used to compute X-ray scatter distributions from CT acquisitions at two or more different spectra simultaneously, in a unified framework with no sacrifice in accuracy and nearly no increase in computation in theory. 
		First, a matrixed-spectrum solver of LBTE is obtained by introducing an additional label dimension to expand the phase space. Then, we propose a ``spectrum basis'' for LBTE and a principle of selection of basis using the QR decomposition, along with the above solver to construct the MSD-LBTE. Based on MSD-LBTE, a unified scatter correction method can be established for multi-spectral CT. 
		We validate the effectiveness and accuracy of our method by comparing it with the Monte Carlo method, including the computational time. We also evaluate the scatter correction performance using two different phantoms for fast-kV switching based dual-energy CT, and using an elliptical phantom in a numerical simulation for kV-modulation enabled CT scans, validating that our proposed method can significantly reduce the computational cost at multiple spectra and effectively reduce scatter artifact in reconstructed CT images.
	\end{abstract}
	\maketitle
	
	\section{Introduction}
	X-ray scatter has been a serious concern since the invention of computed tomography (CT), as it could cause a variety of artifacts in reconstruction and significantly degrade CT image quality if no effective scatter correction or rejection is applied. 
	In the literature, numerous scatter correction or rejection methods have been developed that can be mainly divided into two categories: hardware techniques of scatter rejection and software algorithms of scatter correction. 
	Hardware techniques typically require some sort of interventions in CT system design, including bowtie filter, air-gap and anti-scatter grid\cite{ruhrnschopfGeneralFrameworkReview2011}, which may be less convenient to deploy in clinic.
	
	For software algorithms, statistical methods constitute a category, among which the Monte Carlo (MC) simulation is a classical scattering estimation technique and is considered a gold standard for particle transport simulation\cite{poludniowskiEfficientMonteCarlobased2009,xuPracticalConebeamCT2015}. However, substantial computational requirements and random noise make it quite slowly in order to reach signal levels with low statistic errors, although some acceleration strategies have been developed \cite{gjEfficientScatterDistribution2015,zhangScatterCorrectionBased2020,alsaffarComputationalScatterCorrection2022,sharmaGPUacceleratedFrameworkRapid2021}.
	Recent advancements in deep learning have facilitated the development of learning-based statistical scatter correction methods, and several networks based on supervised and unsupervised learning have been developed \cite{maierDeepScatterEstimation2018,erathDeepLearningbasedForward2021,jiangScatterCorrectionConebeam2019,iskenderScatterCorrectionXRay2022,dongDeepUnsupervisedLearning2021} which perform quite well in scatter correction, but the limited physical interpretability and the needs of training dataset may restrict their applicability in many aspects.
	Another category of scatter correction involves analytic methods. The scatter kernel superposition (SKS) method has been developed in the literature with acceptable accuracy and rapid computational speed\cite{liScatterKernelEstimation2008,maltzAlgorithmXrayScatter2008,sunImprovedScatterCorrection2010}.
	However, the design of scatter kernels that can both achieve generality and accuracy remains challenging, and its efficacy may be limited when applied to complex objects.
	
	It is well known that the linear Boltzmann transport equation (LBTE) can be used to describe the statistical behaviors of photons/particles as they transport in space and interact with matter. In contrast to the MC method, which focuses on individual photon behavior at the microscopic level, the LBTE method tries to solve macroscopic photon distribution throughout the entire spatial domain.
	Scatter correction method based on it has characteristics of high accuracy, good generality, and excellent physical interpretability. Although computationally intensive, the superior parallelism of the equations makes it possible to use GPU and parallel technologies to accelerate the computation process, which has been demonstrated by previous investigations\cite{maslowskiAcurosCTSFast2018,niuUBESUnifiedScatter2024}.
	
	However, for the conventional LBTE based scatter estimation and correction, it is cumbersome if not impossible to calculate the scatter signals at multiple spectra one by one for multi-spectral CT scans such as KV-switching CT and KV modulation CT. As a result, it is highly desired to develop a method that can unitedly compute the scatter signals at multiple spectra in a single calculation with an improved computational efficiency.
	
	In this work, 
	we proposed a novel we propose a Matrixed-Spectrum Decomposition accelerated LBTE solver (MSD-LBTE). 
	First, we derive a semi-analytic solution of the LBTE in a form based on series expansion according to the times of multiple scattering and phase space discretization scheme. Then, we transform the implicit mapping hidden in LBTE into an explicit matrix. 
	After that, we expand the one-dimensional energy group space into two dimensions and obtain a spectrum-matrixed solver of LBTE.
	Based upon these findings, we further bring up a new concept of spectrum basis, and a principle of basis selection using QR decomposition, which makes spectrum basis matrix a lower triangular one and significantly reduces computational redundancy.
	
	In our previous conference paper\cite{zhuUnifiedScatterEstimation2025}, only a preliminary solution was proposed with lack of experimental validations. while in this study, we further optimized the computational efficiency and provided a rigorous mathematical solution formulation. In addition, we have conducted several validations of the computational efficiency and accuracy, and evaluated the feasibility and effectiveness of our proposed method in kV-switching based dual-energy spectral CT and kV-modulation CT scans.
	
	\section{Methods}\label{sec: method}
	\subsection{Linear Boltzmann Transport Equation and Its Numerical Solution}
	\label{sec:LBTE_and_D}
	
	Considering the essential characteristics of photons and the specific requirements of a CT scan, the state of photons can be distinctly defined by their spatial position ($\overrightarrow{r}=(x,y,z)$), flight direction ($\overrightarrow{\Omega}=(\theta,\varphi)$), and energy ($E$), collectively forming the phase space $(\overrightarrow{r},\overrightarrow{\Omega},E)$. The behavior of photons in the phase space during CT scanning can be described by the linear Boltzmann transport equation (LBTE) as follow\cite{maslowskiAcurosCTSFast2018}:
	
	\begin{equation}
		\begin{aligned}
			&\overrightarrow{\Omega} \cdot \nabla \phi(\overrightarrow{r},E,\overrightarrow{\Omega}) + \mu_t(\overrightarrow{r},E)\phi(\overrightarrow{r},E,\overrightarrow{\Omega}) = S(\overrightarrow{r},E,\overrightarrow{\Omega}) \\ +& \int_{E}^{E_{max}} dE' \int_{4\pi}^{} d\overrightarrow{\Omega}' [\mu_s(\overrightarrow{r},E' \rightarrow E,\overrightarrow{\Omega}' \rightarrow \overrightarrow{\Omega})\phi(\overrightarrow{r},E',\overrightarrow{\Omega})].
		\end{aligned}
	\end{equation}
	
	Here, $\phi$ represents the photon fluence, denoting the number of photons passing through a unit section at position $\overrightarrow{r}$ with energy $E$ and flight direction $\overrightarrow{\Omega}$, measured in $m^{-2}sr^{-1}J^{-1}$; $S$ signifies the source term, indicating the number of photons newly generated with energy $E$ and flight direction $\overrightarrow{\Omega}$ in a unit space at position $\overrightarrow{r}$, measured in $m^{-3}sr^{-1}J^{-1}$; $\mu_t$ stands for the linear attenuation coefficient with units of $m^{-1}$; $\mu_s$ denotes the linear scatter coefficient, representing the fraction of photons with energy $E'$ and flight direction $\overrightarrow{\Omega'}$ scattered as photons with energy $E$ and flight direction $\overrightarrow{\Omega}$ in a unit space at position $\overrightarrow{r}$; $E_{max}$ corresponds to the maximum energy of photons. In the field of medical imaging, X-ray energies typically do not exceed 200 keV, well below the pair-production effect whose threshold is 1.022 MeV. Within this energy range, photon-matter interactions are primarily characterized by the photoelectric effect, Compton scattering, and Rayleigh scattering.

	\subsubsection{Semi-Analytic Solution Based on Series Expansion}
	\label{sec:continuous_solution}
	
	As an integral-differential equation, it is often difficult to find an analytic solution for LBTE. To solve such as equation, one way is to classify the photons according to the incident times of multiple scattering, so that source distribution and fluence distribution can be unrolled into series form: $\phi(\overrightarrow{r},E,\overrightarrow{\Omega}) = \sum_{k=0}^{\infty}\phi^{(k)}(\overrightarrow{r},E,\overrightarrow{\Omega})$ and $ S(\overrightarrow{r},E,\overrightarrow{\Omega})= \sum_{k=0}^{\infty}S^{(k)}(\overrightarrow{r},E,\overrightarrow{\Omega})$\cite{maslowskiAcurosCTSFast2018}. When $k = 0$, $S^{(0)}$ represents the X-ray source in a CT scan. After that, the integral-difference equation can be divided into an integral equation and a difference equation, separately.
	
	\begin{subequations}
		\label{series1} 
		\begin{align}
			S^{(k)}(\overrightarrow{r},E,\overrightarrow{\Omega}) = 
			&\overrightarrow{\Omega} \cdot \nabla \phi^{(k)}(\overrightarrow{r},E,\overrightarrow{\Omega}) + \mu_t(\overrightarrow{r},E)\phi^{(k)}(\overrightarrow{r},E,\overrightarrow{\Omega})   \label{seriesa}\\
			S^{(k+1)}(\overrightarrow{r},E,\overrightarrow{\Omega})=   &\int_E^{E_{max}}\int_{4\pi}\mu_s(\overrightarrow{r},E'\rightarrow E,\overrightarrow{\Omega'}\rightarrow \overrightarrow{\Omega})\phi^{(k)}(\overrightarrow{r},E',\overrightarrow{\Omega'})dE'd\overrightarrow{\Omega'},\label{seriesb}
		\end{align}
	\end{subequations}
	Here, Eq.(\ref{seriesb}) shows the explicit transformation from $\phi^{(k)}(\overrightarrow{r},E,\overrightarrow{\Omega})$ to $S^{(k+1)}(\overrightarrow{r},E,\overrightarrow{\Omega})$. It is well know that Eq.(\ref{seriesa}) has an analytic solution as:
	
	\begin{equation}
		\begin{aligned}
			\phi^{(k)}(\overrightarrow{r},E,\overrightarrow{\Omega})&=\int_{0}^{\infty}\frac{S^{(k)}(\overrightarrow{r}-u\overrightarrow{\Omega},E,\overrightarrow{\Omega})}{u^2} \cdot \exp[-\int_0^u u_t(\overrightarrow{r}-v\overrightarrow{\Omega},E)dv]du,
		\end{aligned}
		\label{seriesa2}
	\end{equation}
	where, both $u$ and $v$ represent integral variables. Since $S^{(0)}$ is determined by the geometry of a CT scan and the spectrum of the X-ray source, we can calculate $\phi^{(0)}$ and then $S^{(1)}$ using Eqs.(\ref{seriesa2}) and (\ref{seriesb}), iteratively. Subsequently, by repeatedly  applying the same process, we can determine $\phi^{(1)}$ and $S^{(2)}$ based on the previously calculated $S^{(1)}$. This procedure allows us to obtain $S^{(i)}$ and $\phi^{(i)}$ for all $i \in \mathbf{N}$. The detected signals $I(\overrightarrow{r})$ can be computed as indicated in Eq.(\ref{eq:conti_i}):
	\begin{equation}
		\begin{aligned}
			I(\overrightarrow{r})=\int_0^{E_{max}}dE\int_{4\pi}\phi(\overrightarrow{r},E,\overrightarrow{\Omega})D^{res}(\overrightarrow{r},E)\cdot(\overrightarrow{\Omega}\cdot \overrightarrow{n_d})d\overrightarrow{\Omega},
		\end{aligned}
		\label{eq:conti_i}
	\end{equation}
	where, $\overrightarrow{n_d}$ represents the normal vector of the detector plane, and $D^{res}(\overrightarrow{r},E)$ represents the detector response function. So far, we have derived a series expansion based semi-analytic solution of LBTE, along with an expression for the detector received primary signals and its scatter counterparts. 
	
	\subsubsection{Discretization of Solution}
	\label{sec:discrete_solution}
	
	\begin{figure}[htbp!]
		\centering
		\includegraphics[width=1.0\linewidth]{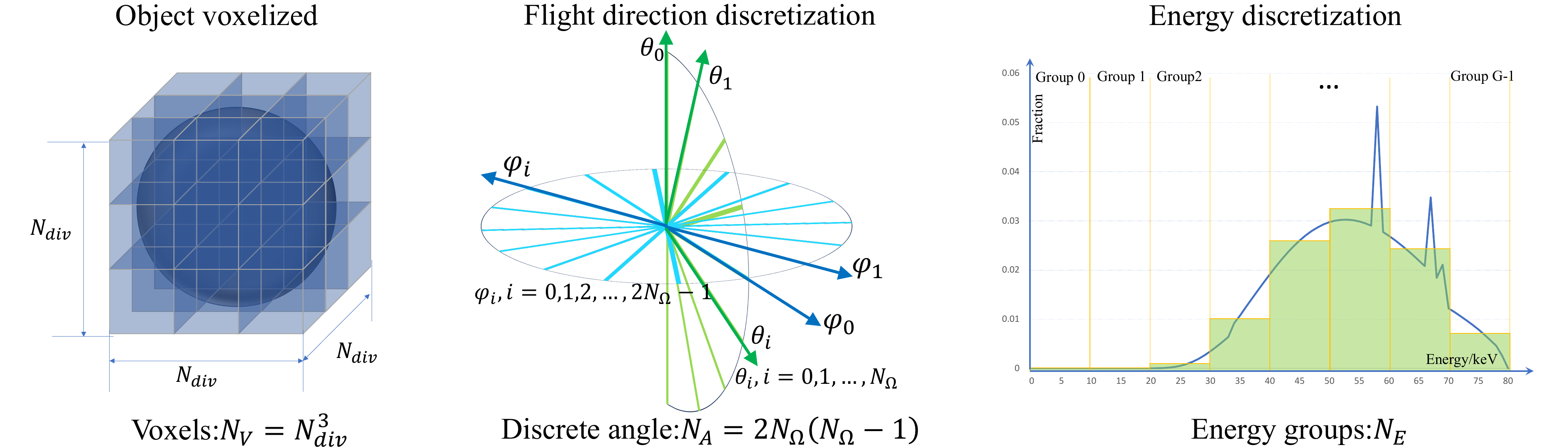}
		\caption{The discretization for phase space $(\overrightarrow{r},E,\overrightarrow{\Omega})$. The object is discretized into $N_{V}$ voxels of independent cubes; the flight direction is discretized based on the distribution of longitude and latitude lines into $N_{A}$ discrete angles; the spectrum is divided into $N_{E}$ groups of independent energies. }
		\label{fig:3_DiscreteScheme}
	\end{figure}
	The LBTE defined in the continuous phase space cannot be directly solved in a digital computing system. Instead, we must discretize the continuous phase space as illustrated in Fig. \ref{fig:3_DiscreteScheme}.
	In reality, the pixelated detectors are naturally discrete. We denote the total number of pixels as $N_{D}$, and use the index $d$ to represent $d$-th pixel.
	In the spatial position domain, we focus mainly on the scanned object and ignore other external factors. One can construct a partition grid within the scanned object, divide it into $N_{V}$ congruent cuboid voxels, and assume that the material composition and photon distribution in a voxel are uniform. Here, index $n$ denotes $n$-th voxel. 
	For energy domains, the continuous spectrum is discretized into a finite number $N_{E}$ of separate energy groups. Suppose that for the $g$-th energy group, the lower bound is $E_g$ and the upper bound is $E_{g+1}$, then the relative intensity $S_g =(\int_{E_g}^{E_{g+1}}S(E)dE)/(E_{g+1} - E_g)$.
	For flight direction domains, we discretized the $4\pi$ solid angle into $M$ discrete angles based on the distribution of longitude and latitude lines. The $(\theta_i,\phi_j)$ coordinates of discrete angle $\overrightarrow{\Omega_m}$ can be written as:
	
	\begin{subequations}
		\label{discrete_angle} 
		\begin{align}
			\theta_i &= i\cdot \frac{\pi}{N_{\Omega}},i=0,1,2,...,N_{\theta}-1, N_{\theta},\\
			\varphi_j &= j \cdot \frac{2\pi}{N_{\Omega}}, j=0,1,2,...,N_{\varphi}-2,N_{\varphi}-1.
		\end{align}
	\end{subequations}
	When $i=0$ or $N_{\Omega}$, the discrete angle is equal to the "pole" in longitude and latitude, and for different $j$ it represents the same direction. Here, the values of $N_{A}$ and $N_{\Omega}$ are associated as $N_{A} = 2N_{\Omega}(N_{\Omega} -1) +2$. Based on the above discretization scheme, we divide the continuous phase space $(\overrightarrow{r},E,\overrightarrow{\Omega})$ into the discrete phase space $(\overrightarrow{r_n},E_g,\overrightarrow{\Omega_m})$.
	
	In the discrete phase space, the discrete forms of Eqs.(\ref{seriesa2}) and (\ref{seriesb}) can be rewritten as:
	\begin{subequations}
		\label{discrete_lbte_trans_and_Scat}
		\begin{align}
			\phi^{(k)}_{n,g,m_{n'\rightarrow n}} &= \sum_{n'=0}^{N_{V}-1} \frac{\exp(-\sum_{w'}^{}\mu_{t;n_{w'},g}\cdot  w')}{|\overrightarrow{r_n} - \overrightarrow{r_{n'}}|^2} S^{(k)}_{n',g,m_{n'\rightarrow n}}, k\geq 1, \label{discrete_lbte_trans_and_Scat_a}\\
			S^{(k+1)}_{n,g,m} &= \sum_{g'=0}^{N_{E}-1} \sum_{m'=0}^{N
				_{a}-1} \mu_{s;n,g' \rightarrow g, m' \rightarrow m}\phi^{(k)}_{n,g',m'},k\geq 0 ,\label{discrete_lbte_trans_and_Scat_b}
		\end{align}
	\end{subequations}
	where, $w'$ represents the path-length in voxel $i$, $\mu_{t:i,g}$ represents the average linear attenuation coefficient in voxel $i$ in energy group $g$. For voxel $i$ containing more than one material, $\mu_{t:i,g} = \sum_{n=1}^{N}\mu_{m:n,g} \rho_n$, where $\mu_{m:n,g}$ represents the average mass attenuation coefficient of material $n$ in energy group $g$ and $\rho_n$ represents the density of material $n$ in voxel $i$.

	In particular, $S^{(0)}(\overrightarrow{r},E,\overrightarrow{\Omega})$ differs from $S^{(k)}(\overrightarrow{r},E,\overrightarrow{\Omega}),k\geq 1$, because $S^{(0)}$ represents the zero-order X-ray source in a CT scan rather than the higher-order X-ray source in the voxel. Therefore, the spatial position of $S^{(0)}$ cannot be represented by discrete voxel positions because $S^{(0)}$ is outside the scanned object. In continuous phase space, the expression of $S^{(0)}(\overrightarrow{r},E,\overrightarrow{\Omega})$ is given by  $S^{(0)}(\overrightarrow{r},E,\overrightarrow{\Omega}) = O(\overrightarrow{\Omega})S(E)\delta(\overrightarrow{r}-\overrightarrow{r_s})$, where $\overrightarrow{r_s}$ represents the spatial location of the X-ray source, $S(E)$ represents the spectrum of X-ray source and $O(\overrightarrow{\Omega})$ represents the focus range of the X-ray source. After discretization, it is seen that $S^{(0)}_{g,m}=O_{m}\cdot S_g$. In the case of $k=0$, Eq.(\ref{discrete_lbte_trans_and_Scat_a}) is simplified as:
	\begin{equation}
		\begin{aligned}
			\label{phi0_to_Sg}
			\phi^{(0)}_{n,g,m_{\overrightarrow{r_s} \rightarrow \overrightarrow{r_{n}}}} &= \frac{\exp(-\sum_{w'}^{}\mu_{t;n_{w'},g}\cdot  w')}{|\overrightarrow{r_s} - \overrightarrow{r_{n}}|^2} O_{m_{\overrightarrow{r_s} \rightarrow \overrightarrow{r_{n}}}}\cdot S_g,
		\end{aligned}
	\end{equation}
	
	Finally, the discrete forms of the primary and scatter signals received by the detector can be written as:
	
	\begin{subequations}
		\label{Ip_and_Is}
		\begin{align}
			I^S_{d}&=\sum_{n=0}^{N_{V}-1} \sum_{g=0}^{N_{E}-1}  \frac{a_d \cos(\overrightarrow{\Omega_m}\cdot \overrightarrow{n_d}) E_{g'} D^{res}_{d,g}\cdot }{|\overrightarrow{r_n}-\overrightarrow{r_d}|^2} \cdot \exp(-\sum_{l'}^{}\mu_{t;i_{l'},g} \cdot w') \sum_{k=1}^{N_{S}}S^{(k)}_{i,g,m_{\overrightarrow{r_i}\rightarrow \overrightarrow{r_d}}},\label{Is}\\
			I^P_{d} &= \sum_{g=0}^{N_{E}-1}\frac{a_d \cos(\overrightarrow{\Omega_m}\cdot \overrightarrow{n_d}) E_{g'} D^{res}_{d,g}}{|\overrightarrow{r_s}-\overrightarrow{r_d}|^2}\cdot \exp(-\sum_{l'}^{}\mu_{t;i_{l'},g} \cdot w')O_{m_{\overrightarrow{r_s} \rightarrow \overrightarrow{r_{d}}}}\cdot S_g, \label{Ip}
		\end{align}
	\end{subequations}
	where, $\overrightarrow{r_d}$ represents the spatial position of pixel $d$, $a_d$ represents the area of pixel $d$, $D^{res}_{d,g}$ represents the discrete form of $D^{res}(\overrightarrow{r},E)$, and $N_{S}$ represents the upper limit of scattering. In practice, the number of photons in a higher-order of multiple scattering is smaller and their contributions to the overall scattered photons is mostly negligible, so one can truncate at a finite order of multiple scattering. So far, we have turned formulas defined in continuous phase space into discrete forms, which can be solved in a computer.

	\subsection{Spectrum-Matrixed Solver for LBTE and a Concept of Spectrum Basis}
	\label{sec:MSD-LBTE}
	
	\begin{figure}[htbp!]
		\centering
		\includegraphics[width=0.9\linewidth]{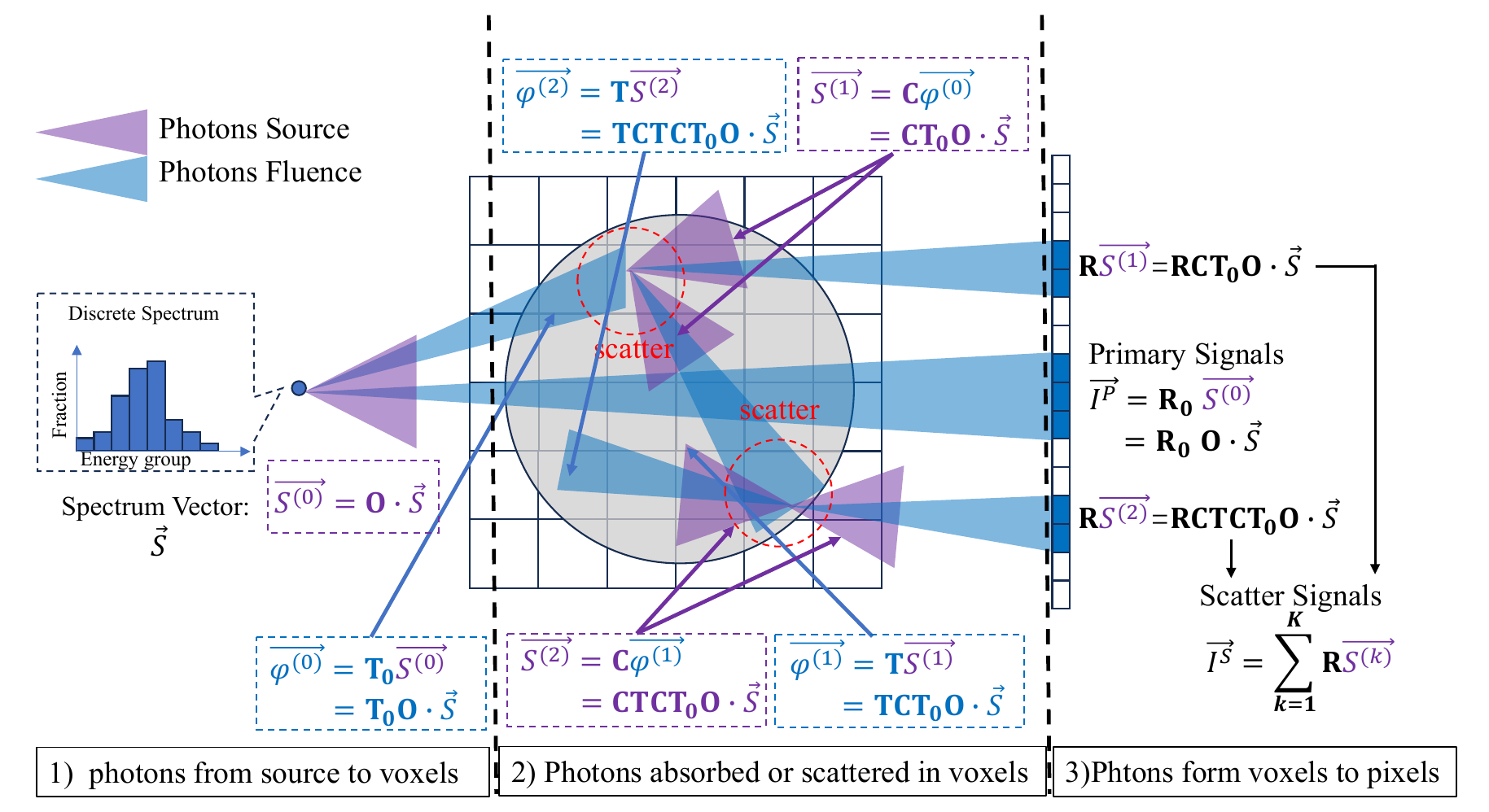}
		\caption{The matrixied form of the entire LBTE numerical expressions}
		\label{fig:4_LbteIteration}
	\end{figure}
	
	For ease of narration, we convert the above discrete expressions into matrix form. Consider $\phi^{(k)}_{n,g,m},k\geq0$ and $S^{(k)}_{n,g,m},k\geq1$ as ${N_{V}\times N_{E}\times N_{A}}$-dimension vectors, respectively, denoted by $\overrightarrow{\phi^{(k)}}$ and $\overrightarrow{S^{(k)}}$, where $\overrightarrow{\phi^{(k)}},\overrightarrow{\phi^{(k)}} \in \mathbb{R}^{N_{V}N_{E}N_{A}}$. Similarly, $I^S_d$, $I^P_d$ are represented by vectors $\overrightarrow{I^S}$, $\overrightarrow{I^P}$ $\in \mathbb{R}^{N_{D}}$. and $S_g$ is represented by the vector $\overrightarrow{S} \in \mathbb{R}^{N_{E}}$. Then the Eqs.(\ref{discrete_lbte_trans_and_Scat})-(\ref{Ip_and_Is}) can be rewritten to the following matrix forms:
	
	\begin{subequations}
		\label{matrixform}
		\begin{align}
			\overrightarrow{\phi^{(k)}} &= \mathbf{T}\overrightarrow{S^{(k)}}, k\geq 1;\overrightarrow{S^{(k+1)}} =\mathbf{C}\overrightarrow{\phi}^{(k)},k\geq 0,\label{matrixform_a}\\
			\overrightarrow{\phi^{(0)}} &= \mathbf{T_0}\mathbf{O}\overrightarrow{S},\label{matrixform_b}\\
			\overrightarrow{I^S} &= \mathbf{R}\sum_{k=1}^{N_{S}}\overrightarrow{S^{(k)}},\label{matrixform_c}\\
			\overrightarrow{I^P} &= \mathbf{R_0}\mathbf{O}\overrightarrow{S},\label{matrixform_d}
		\end{align}
	\end{subequations}
	combining Eqs.((\ref{matrixform_a}), (\ref{matrixform_b}), and (\ref{matrixform_c}), one gets: 
	\begin{equation}
		\label{matrixform_f}
		\begin{aligned}
			\overrightarrow{I^S} &= \mathbf{R}\left(\sum_{k=1}^{K}\mathbf{(TC)}^{k-1}\right)\mathbf{C}\mathbf{T_0}\mathbf{O}\cdot \overrightarrow{S}.
		\end{aligned}
	\end{equation}
	
	Here,
	
	$\mathbf{T_0} \in \mathbb{R}^{N_{V}N_{E}N_{A}\times N_{E}N_{A}}$ is the primary transport matrix, and $\mathbf{R_0} \in \mathbb{R}^{N_{D}\times N_{E}N_{A}}$ is primary detector matrix. Both of them are sparse matrices determined by the source position and material density distribution. $\mathbf{R_0}$ also depends on the detector position and response function.

	$\mathbf{O} \in \mathbb{R}^{N_{E}N_{A}\times N_{E}}$ is primary X-ray source focus range matrix, which determined by X-ray source position and its field of view (FOV).

	$\mathbf{T} \in \mathbb{R}^{N_{V}N_{E}N_{A}\times N_{V}N_{E}N_{A}}$ is the secondary transport matrix, $\mathbf{C} \in \mathbb{R}^{N_{V}N_{E}N_{A}\times N_{V}N_{E}N_{A}}$ is secondary scatter matrix, and $\mathbf{R} \in \mathbb{R}^{N_D\times N_{V}N_{E}N_{A}}$ is secondary detector matrix. All of them are sparse matrices determined by material density distribution. $\mathbf{R}$ also depends on the detector position and response function.

	\subsubsection{Phase Space Expansion and Spectrum-Matrixing}
	
	In order to fully utilize the parallelism of LBTE, we introduce a label dimension ($l$) to the phase space, which assigns an additional information to the photon emitted from the X-ray source. As a result, the phase space expands from $(\overrightarrow{r_k}, E_g, \overrightarrow{\Omega_m})$ to $(\overrightarrow{r_n}, E_g, \overrightarrow{\Omega_m}, l)$, $l \in \{0,1,...,N_{L}-1\}$. 
	Then, $S^{(k)},\phi^{(k)}$ in Eq.(\ref{matrixform}) are expanded as $\overrightarrow{S^{(k)}_l},\overrightarrow{ \phi^{(k)}_l},l=0,1,2,...,N_{L}-1$, $\overrightarrow{I^S}$, $\overrightarrow{I^P}$ are expanded as $\overrightarrow{I^S_l},\overrightarrow{I^P_l},l=0,1,2,...,N_{L}-1$, $\overrightarrow{S}$ is expanded as $\overrightarrow{S_l},l=0,1,2,...,N_{L}-1$, and $\mathbf{T},\mathbf{T_0},\mathbf{R},\mathbf{R_0},\mathbf{C}$ and $\mathbf{O}$ will not be expanded because they are independent of the information of photons and instead only dependent of geometry, detector response, X-ray source FOV or material density distribution. After such as phase space expansion, Eqs.(\ref{matrixform_d}) and (\ref{matrixform_f}) can be rewritten as:
	
	\begin{subequations}
		\label{matrixform_expanded}
		\begin{align}
			\overrightarrow{I_l^S} &= \mathbf{R}\left(\sum_{k=1}^{N_{S}}\mathbf{(TC)}^{k-1}\right)\mathbf{C}\mathbf{T_0}\mathbf{O}\cdot \overrightarrow{S_l}&,l=0,1,2,...,N_{L}-1,\label{matrixform_expanded_s}\\
			\overrightarrow{I_l^P} &= \mathbf{R_0}\mathbf{O}\overrightarrow{S_l}&,l=0,1,2,...,N_{L}-1,\label{matrixform_expanded_p}
		\end{align}
	\end{subequations}
	and a series of spectrum vectors $\overrightarrow{S_l},l=0,1,2,..,N_{L}-1$ can be written as a single spectrum matrix $\mathbf{S}=[\overrightarrow{S_0},\overrightarrow{S_1},...,\overrightarrow{S_{N_{L}-1}}]$, where each column vector of spectrum matrix $\mathbf{S}$ represents a spectrum vector $\overrightarrow{S}$ under different $l$. As such a process means that the spectrum space changes from $\mathbb{R}^{N_{E}}$ to $\mathbb{R}^{N_{E}\times N_{L}}$, it is called ``spectra-matrixing'' in this paper. 
	
	Similarly, $\overrightarrow{I_l},\overrightarrow{I_s},l=0,1,2,..,N_{L}-1$ can also be written as matrix $\mathbf{I^S}=[\overrightarrow{I^S_0},\overrightarrow{I^S_1},...,\overrightarrow{I^S_{N_{L}-1}}], \mathbf{I^P}=[\overrightarrow{I^P_0},\overrightarrow{I^P_1},...,\overrightarrow{I^P_{N_{L}-1}}]$ and the detected signal space changes from $\mathbb{R}^{N_{D}}$ to $\mathbb{R}^{N_{D}\times N_{L}}$. The same applies to $\overrightarrow{\phi^{(k)}}$ and $\overrightarrow{S^{(k)}}$, as they will be expanded into matrices $\mathbf{\Phi^{(k)}}$ and $\mathbf{S^{(k)}}$ with $\mathbb{R}^{N_{V}N_{E}N_{A}}$ expanded into $\mathbb{R}^{N_{V}N_{E}N_{A}\times N_{L}}$.
	
	If we set $\mathbf{A^s} = \mathbf{R}\left(\sum_{k=1}^{N_{S}}\mathbf{(TC)}^{k-1}\right)\mathbf{C}\mathbf{T_0}\mathbf{O}$ and $\mathbf{A^p} = \mathbf{R_0O}$, then Eqs.(\ref{matrixform_d}) and (\ref{matrixform_f}) become:
	
	\begin{subequations}
		\begin{align}
			\mathbf{I^p} &=\mathbf{A^p} \mathbf{S},\\
			\mathbf{I^s} &= \mathbf{A^s} \mathbf{S}
		\end{align}
	\end{subequations}
	
	In such a way above, the mappings from the spectrum to primary and scatter signals ,implied in LBTE ,are represented explicitly as matrices $\mathbf{A^p}$ and $\mathbf{A^s}$. In order to facilitate the narration, we use $\mathbf{A}\in \mathbb{R}^{N_{D}\times N_{E}}$ to represent these two matrices, and use $\mathbf{I}\in \mathbb{R}^{N_{D}\times N_{L}}$ to represent $\mathbf{I^P}$ and $\mathbf{I^S}$, where $\mathbf{S} \in \mathbb{R}^{N_{E}\times N_{L}}$. 
	
	Here, it is worth noting that the introduction of label dimension adds just a small computational overhead. When $N_{L}=1$, let the computation cost be 1 unit. When $N_{L}=2$, the computation cost does not increase to 2 unit but to $1+t$ unit with $t <1$ (for $\forall N_{L}$, the computation cost equals $1+(N_{L}-1)t$ unit). Because $\mathbf{A}$ are independent of dimension $l$, we only need to calculate them once. Adding $N_{L}$ labels only increases about $2N_{S}(N_{L}-1)$ times of sparse matrix multiplication as $(N_{L}-1)t$ unit computation cost (refer to Eq.(\ref{matrixform})), which is relatively small when compared with the cost of computing these sparse matrices. 
	The subsequent experiments will validate this assumption. 
	So far, we have established a matrixed-spectrum solver for LBTE (MS-LBTE).

	\subsubsection{A Novel Concept with Spectrum Basis Vectors}
	\label{sec:motivation}
	
	In order to obtain different scatter signals for multiple spectra in a single computation, we need to optimize the process of LBTE solving and reduce the computational redundancy.
	Fortunately, we notice that process of solving LBTE is actually to find a linear implicit mapping from the spectrum $S(E)$ to the detected signal $I(\vec{r})$ as $f:S(E) \rightarrow I(\overrightarrow{r})$.
	After discretization as shown in Sec. \ref{sec:LBTE_and_D}, the continuous variable $S$ becomes an $N_{E}$-dimensional vector $\overrightarrow{S}$, and $I$ becomes $N_{D}$-dimensional vectors $\overrightarrow{I}$. The following relationship still holds: $\overrightarrow{I} = f(\overrightarrow{S})$.

	Since $\overrightarrow{S}$ is in $\mathbb{R}^G$, it is easy to pick a set of linearly independent $\overrightarrow{S^{basis}_0},\overrightarrow{S^{basis}_1},...\overrightarrow{S^{basis}_{G-1}}$ as the spectrum basis vectors in $\mathbb{R}^G$. Then, for $\forall \overrightarrow{S} \in \mathbb{R}^G$, we can get a set of coefficients $c^{basis}_0,c^{basis}_1,...,c^{basis}_{N_{E}-1}$ that satisfy:
	\begin{equation}
		\begin{aligned}
			\overrightarrow{S} = c^{basis}_0\overrightarrow{S^{basis}_0} + c^{basis}_1\overrightarrow{S^{basis}_1} + ...+ c^{basis}_{N_{E}-1}\overrightarrow{S^{basis}_{N_{E}-1}}
		\end{aligned}
		\label{eq:basis_series}
	\end{equation}
	Given that $f$ is a linear mapping, we have:
	
	\begin{equation}
		\begin{aligned}
			\overrightarrow{I} 
			& = f(\overrightarrow{S}) \\
			& = f(c^{basis}_0\overrightarrow{S^{basis}_0} + c^{basis}_1\overrightarrow{S^{basis}_1} + ...+ c^{basis}_{N_{E}-1}\overrightarrow{S^{basis}_{N_{E}-1}}) \\
			& = c^{basis}_0 f(\overrightarrow{S^{basis}_0}) + c^{basis}_1 f(\overrightarrow{S^{basis}_1}) + .. + c^{basis}_{N_{E}-1} f(\overrightarrow{S^{basis}_{N_{E}-1}}) \\
			& = c^{basis}_0\overrightarrow{I^{basis}_0} + c^{basis}_1 \overrightarrow{I^{basis}_1} + ... +c^{basis}_{N_{E}-1} \overrightarrow{I^{basis}_{N_{E}-1}}
		\end{aligned}
		\label{eq:basis_compos}
	\end{equation}
	where,
	\begin{equation}
		\begin{aligned}
			\overrightarrow{I^{basis}_i} = f(\overrightarrow{S^{basis}_i}),i=0,1,...,N_{E}-1
		\end{aligned}
		\label{eq:basis_trans}
	\end{equation}
	with $\overrightarrow{I^{basis}_i},i=0,1,...,N_{E}-1$ being detected signal basis vectors.

	By using such spectrum bases, for any spectrum $\overrightarrow{S}$, we do not need to calculate the corresponding $\overrightarrow{I}$ one by one. Instead, we can first select a set of spectrum basis vectors and obtain a set of coefficients that $\overrightarrow{S}$ resolves under that bases as in Eq.(\ref{eq:basis_series}). Then we calculate a set of detected signals basis vectors based on spectrum basis vectors as in Eq.(\ref{eq:basis_trans}). Finally, we can get the detected signal $\overrightarrow{I}$ corresponding to $\overrightarrow{S}$, which is equal to a linear combination of detected signal bases as in Eq.(\ref{eq:basis_compos}).

	An obvious advantage of such a scheme above is that no matter how many spectra ($N_{spec}$) we used (even $N_{spec}>N_{E}$), we only need to compute $N_{E}$ detected signal vectors $\overrightarrow{I^{basis}_i},i=0,1,...,N_{E}-1$ based on G selected spectrum basis vectors $\overrightarrow{S^{basis}_i},i=0,1,...,N_{E}-1$ as in Eq.(\ref{eq:basis_trans}). In addition, if we only use $N_{spec}$ different spectra $\overrightarrow{S_i}, i=0,1,...,N_{spec}-1$ and $N_{spec} < N_{E}$, these vectors cannot fill the whole $\mathbb{R}^{N_{E}}$ space. In fact, the ${N_{spec}}$ spectra vectors span only the lower dimensional $\mathbb{R}^{N_{spec}}$ space ($\mathbb{R}^{N_{spec}}\subset \mathbb{R}^{N_{E}}$) in fact. As a result, we can select fewer spectrum basis vectors $\overrightarrow{S^{basis}_i},i=0,1,...,{N_{spec}}-1$, leading to fewer computations of detected signal basis vectors $\overrightarrow{I^{basis}_i},i=0,1,...,{N_{spec}}-1$.
	
	In MS-LBTE, the implicit mapping $f$ is transformed into an explicit matrix $\mathbf{A} \in \mathbb{R}^{N_{D}\times N_{E}}$, and spectrum vector $\overrightarrow{S} \in \mathbb{R}^{N_{E}}$ and detected signal vector $\overrightarrow{I} \in \mathbb{R}^{N_{D}}$ are expanded into $\mathbb{R}^{N_{E}\times N_{L}}$ and $\mathbb{R}^{N_{E}\times N_{L}}$. With no spectrum basis strategy above, if there are $N_{spec}$ different spectra, it is considered that $l$ marks different spectrum vectors used so that $N_{L} = N_{spec}$, which can be written as $\mathbf{I} = \mathbf{A} \cdot \mathbf{S}$, with $\mathbf{I}=[\overrightarrow{I_0},\overrightarrow{I_1},...,\overrightarrow{I_{N_{spec}-1}}]\in \mathbb{R}^{N_{D}\times N_{spec}}$ and $\mathbf{S}=[\overrightarrow{S_0},\overrightarrow{S_1},...,\overrightarrow{S_{N_{spec}-1}}]\in \mathbb{R}^{N_{E}\times N_{spec}}$.
	
	Using the spectrum basis strategy, we can write multiple spectrum basis vectors as a matrix $\mathbf{S^{basis}} = [\overrightarrow{S^{basis}_0},\overrightarrow{S^{basis}_1},\overrightarrow{S^{basis}_2},...]$. Similarly we have $\mathbf{I^{basis}}=[\overrightarrow{I^{basis}_0},\overrightarrow{I^{basis}_1},\overrightarrow{I^{basis}_2},...]$ and $\overrightarrow{C^{basis}}=[c^{basis}_0;c^{basis}_1;c^{basis}_2;...]$, then we can obtain the basis decomposition of $\overrightarrow{S}$ as $\overrightarrow{S} = \mathbf{S^{basis}}\overrightarrow{C^{basis}}$ corresponding to Equation (\ref{eq:basis_series}) and the linear combination of $\overrightarrow{I}$ as $\overrightarrow{I} = \mathbf{I^{basis}}\overrightarrow{C^{basis}}$ corresponding to Eq.(\ref{eq:basis_compos}).

	If there are $N_{spec}$ different spectra, now $\overrightarrow{I_i}=\mathbf{A}\overrightarrow{S_i},i=0,1,...,N_{spec}-1$. Let's $\mathbf{I}=[\overrightarrow{I_0},\overrightarrow{I_1},...,\overrightarrow{I_{N_{spec}-1}}]$ ($\mathbf{I} \in \mathbb{R}^{N_{D}\times N_{spec}}$)
	and $\mathbf{S}=[\overrightarrow{S_0},\overrightarrow{S_1},...,\overrightarrow{S_{N_{spec}-1}}]$ ($\mathbf{S} \in \mathbb{R}^{N_{E}\times N_{spec}}$), we can obtain $\mathbf{S}=\mathbf{S^{basis}C^{basis}}$ and $\mathbf{I}=\mathbf{I^{basis}C^{basis}}$, with $\mathbf{C^{basis}}=[\overrightarrow{C^{basis}_0},\overrightarrow{C^{basis}_1},\overrightarrow{C^{basis}_2},...]$ and $\overrightarrow{S_i}=\mathbf{S} \overrightarrow{C^{basis}_i}$. Then we can finally obtain the following expressions,
	
	\begin{subequations}
		\label{AS_equal_I_vector}
		\begin{align}
			\mathbf{S} &= \mathbf{S^{basis}}\mathbf{C^{basis}},\label{AS_equal_I_vector_a}\\
			\mathbf{I^{basis}} &= \mathbf{A}  \mathbf{S^{basis}} ,\label{AS_equal_I_vector_b}\\
			\mathbf{I} &= \mathbf{I^{basis}}\mathbf{C^{basis}},\label{AS_equal_I_vector_c} 
		\end{align}
	\end{subequations}
	
	Such an introduction of spectrum basis matrix serves as a transitional position, which enable us to freely select the spectrum matrix used in LBTE solving without being restricted by the actual spectrum matrix determined by X-ray source as in the following Sec.\ref{sec:spectramatrixed}.
	
	\subsection{Spectrum Basis Optimization with QR Decomposition }
	\label{sec:spectramatrixed}
	
	\begin{figure}[htbp!]
		\centering
		\includegraphics[width=1.0\linewidth]{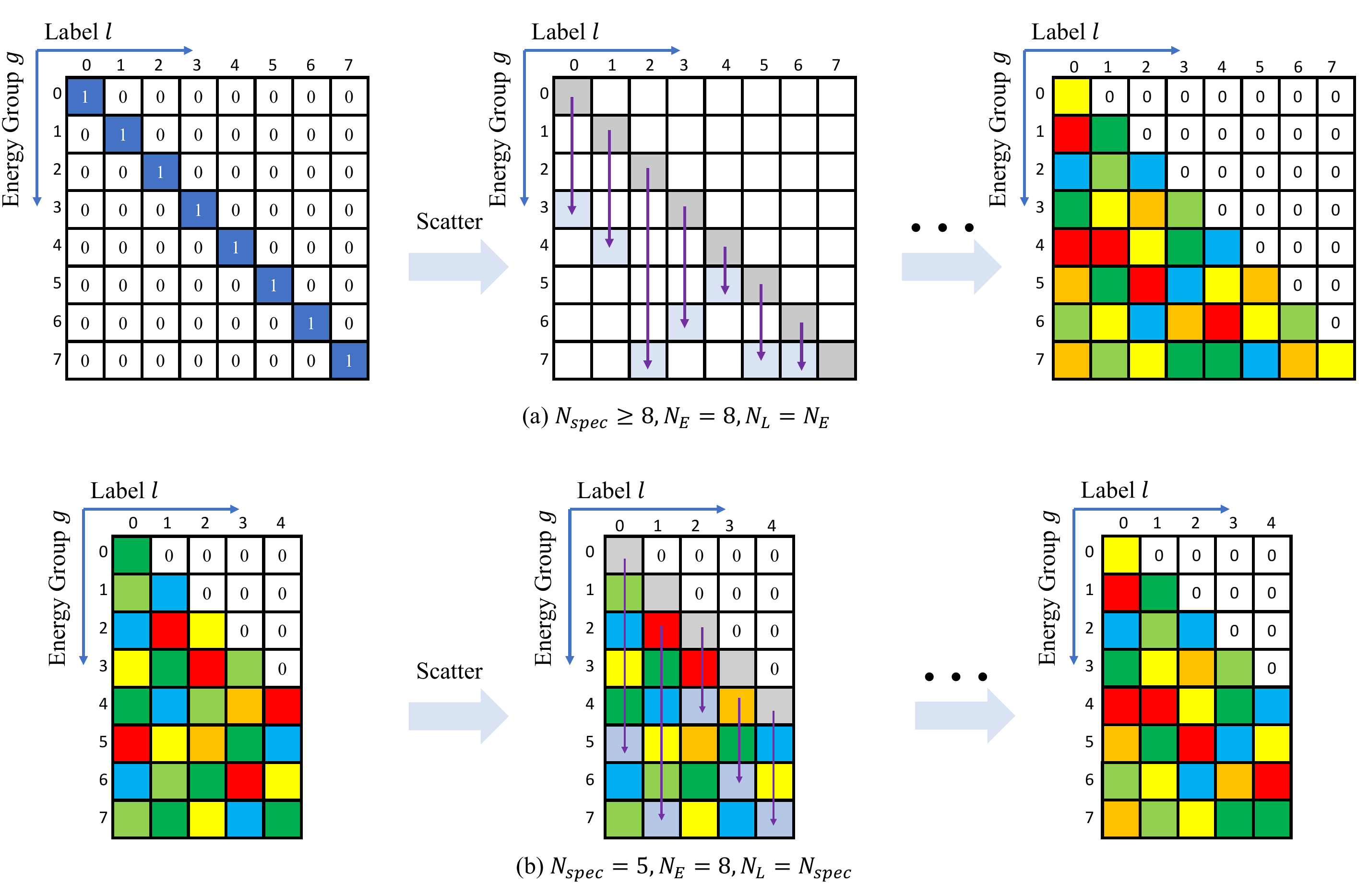}
		\caption{
			The proportion of photons in each energy group for each basis spectrum, which are determined by the spectrum basis matrix (take $N_E=8$ for example). Different colors represent different proportions, while white represents a proportion of zero. (a) $N_{spec} \geq 8$ and $N_{L}=\min(N_{E},N_{spec})=8$ is selected, where the spectrum basis matrix is an identity matrix, with only $l = g$ having a nonzero value of $1$ initially; (b) $N_{spec} = 5$ and $N_{L}=\min(N_{E},N_{spec})=5$ is selected, where the spectrum basis matrix is a lower triangular matrix, with only $l \leq g$ having a nonzero values initially. When scattering happens, a high energy group of photons will migrate into a low energy group. In the whole process, the upper triangular portion ($l > g$) of the matrix is always zero during scattering, meaning no computation is needed for those energy groups of photons, so the computation can be reduced.}
		\label{fig:5_QRdecomposition}
	\end{figure}
	
	When $N_{spec} \geq N_{E}$, we can set $N_{L}=N_{E}$. The spectrum matrix space is $\mathbb{R}^{N_{E} \times N_{E}}$. We select a set of unit orthogonal spectrum basis vectors $\overrightarrow{e_0},\overrightarrow{e_1},...,\overrightarrow{e_{N_{E}-1}}$ in $\mathbb{R}^G$, where $\overrightarrow{e_0}=[1,0,...,0],\overrightarrow{e_1}=[0,1,...,0],...,\overrightarrow{e_{N_{E}-1}}=[0,0,..,1]$. Then,$\mathbf{S^{basis}}=[\overrightarrow{e_0},\overrightarrow{e_1},...,\overrightarrow{e_{N_{E}-1}}] = \mathbf{E}$, which is an identity matrix in $\mathbb{R}^{N_{E}\times N_{E}}$. In theory, the computational cost is $1 + (N_{E}-1)t$ in this case. Under such circumstances, as the photoelectric effect, Rayleigh scatter and Compton scattering effect are considered in the equation, the energy of photons will merely decrease rather gradually than increase. Let energy group $0$ be the highest energy group and energy group $G-1$ be the lowest energy group. Then, for the spectrum basis vectors $\overrightarrow{e_l}$ (i.e., the energy spectrum where only energy group $l$ has a non-zero value), only the behavior of photons in the lower energy groups $l, l + 1, l + 2,..., N_{E} - 1$ needs to be taken into account, while the behavior of the higher energy groups $1, 2,..., l - 1$ need no consideration as shown in Fig. \ref{fig:5_QRdecomposition}(a). This enables the computational overhead to be further reduced, from $1+(N_{E}-1)t$ unit to $1 + (N_{E}-1)t\cdot \frac{N_{E}+1}{2N_{E}}) $ unit instead.

	When $N_{spec} < N_{E}$, we set $N_{L}=N_{spec}$, the spectrum matrix space is $\mathbb{R}^{N_{E}\times N_{spec}}$. We can also select a set of spectrum basis vectors $\overrightarrow{b_0},\overrightarrow{b_1},...,\overrightarrow{b_{N_{E}-1}}$ which form a lower triangular matrix $\mathbf{B}$ as shown in Fig. \ref{fig:5_QRdecomposition}(b). 
	
	Luckily, the validity of such a selection can be guaranteed by the QR decomposition of a matrix. In other words, We can obtain $\mathbf{S^{basis}}$ and $\mathbf{C^{basis}}$ by finding the QR decomposition of the transpose of the matrix $\mathbf{S}$ as,
	\begin{equation}
		\begin{aligned}
			\mathbf{S^T}=\mathbf{Q_s R_s}
		\end{aligned}
	\end{equation}
	where $\mathbf{Q_s}$ is an orthogonal of size $N_{spec}\times N_{spec}$ and $\mathbf{R_s}$ is an upper triangular matrix. Taking the transpose of both sides, we obtain, 
	\begin{equation}
		\begin{aligned}
			\mathbf{S}=\mathbf{R_s^T}\mathbf{Q_s^T}
		\end{aligned}
	\end{equation}
	Then, compared with Eq.(\ref{AS_equal_I_vector_a}), we can get $\mathbf{S^{basis}}=\mathbf{R_s^T},\mathbf{C^{basis}}=\mathbf{Q_s^T}$. 
	
	Now, $\mathbf{S^{basis}}$ is a lower-triangular matrix. Considering the differences between different spectra, we can think of $\mathbf{S}$ as a column full rank matrix, which avoids column pivoting in the QR decomposition. It is worth noting that the $\mathbf{S^{basis}}$ obtained by the QR decomposition may contain negative values which has no mean in physics, but will have no negative impact on final outcome in computation as our LBTE solving process is linear with respect to the number of photon at a specific energy level. In such a case, the computational cost is no longer $1+(N_{spec}-1)t$ unit. It becomes $1 + (N_{spec}-1)t\cdot( 1 - \frac{N_{spec} - 1}{2N_{E}}) $ unit instead, leading to an LBTE solver with better computational efficiency. In this paper, such a matrixed-spectrum decomposition accelerated solver of LBTE is called MSD-LBTE.
	
	\subsection{Unified and Accelerated Scatter Correction in Multi-Spectral CT}
	\begin{figure}[htbp!]
		\centering
		\includegraphics[width=1.0\linewidth]{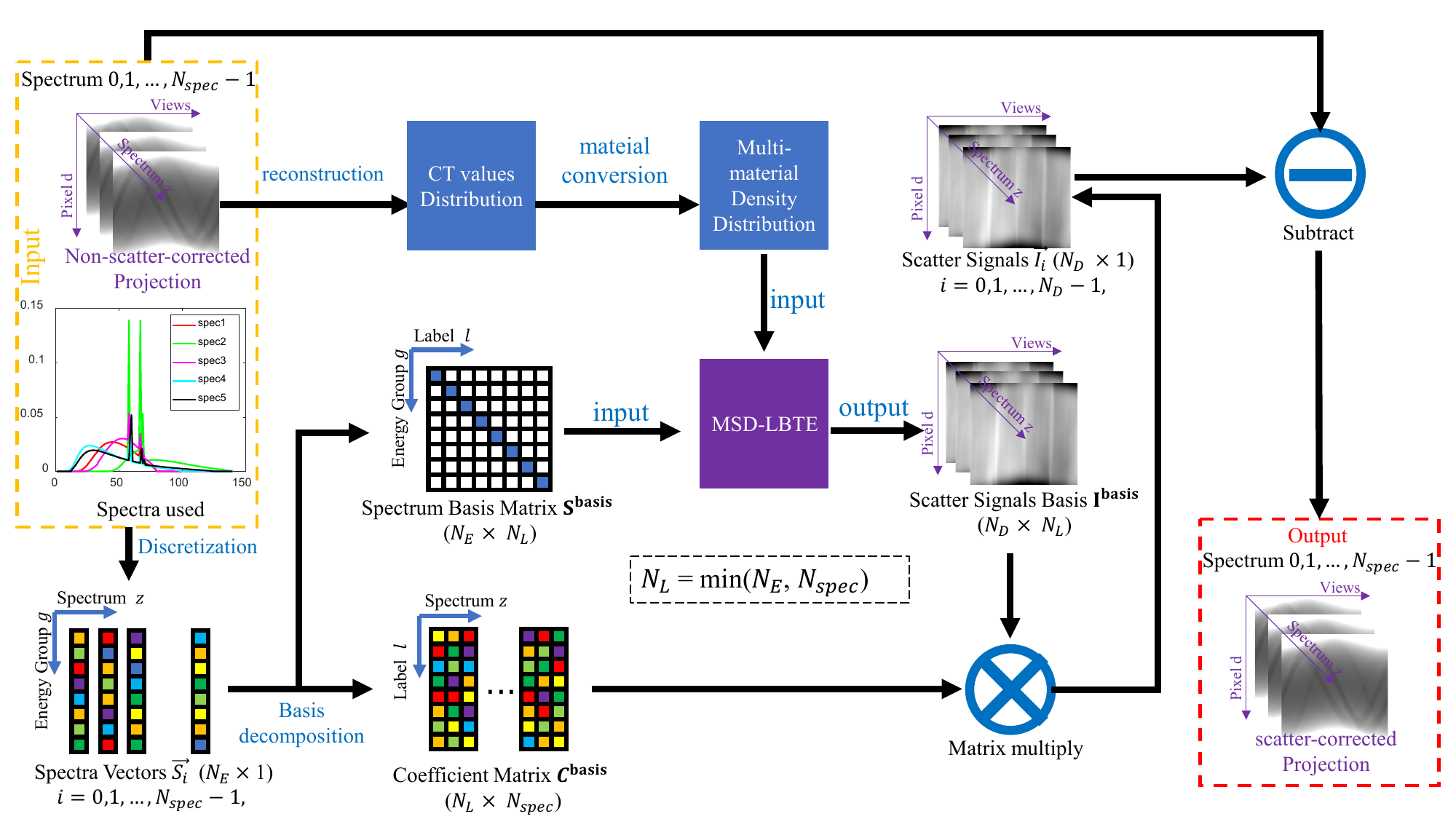}
		\caption{A unified scatter correction method based on our proposed MSD-LBTE method.}
		\label{fig:6_Pipeline}
	\end{figure}
	
	Based on the above deviations and findings, a unified scatter correction for multi-spectral CT can be straightforwardly implemented, as illustrated in Fig. \ref{fig:6_Pipeline}. 
	Multiple steps are usually involved. First, it is essential to preprocess the projection data to generate the initial CT images. Subsequently, the distributions of CT values need to be converted into material densities, serving as inputs of scanned objects for MSD-LBTE to.
	After that, the spectrum matrix, consisting of multiple spectrum vectors, is decomposed into a product of the spectrum basis matrix and a coefficient matrix. The spectrum basis matrix, along with the multi-material densities of the object, is fed into MSD-LBTE to generate the scatter signal bases. The bases are then multiplied by the coefficient matrix to derive the actual scatter signals needed for each specific spectrum.
	Ultimately, a set of scatter signals are obtained, usually with an appropriate scaling factor for scatter correction determined in advance practically. In the end, the scatter signals are then subtracted from the total transmission data to generate scatter-corrected projections from which final reconstructions can be done.
	In comparison with the conventional LBTE based ones, our method offers a clear advantage of simultaneously scatter computation for multiple spectra, leading to a significantly improved computational efficiency, with no sacrifice in computational accuracy.
	
	\subsection{Experimental Setups}
	
	In order to validate our proposed MSD-LBTE method, we measured the computational time of using conventional LBTE methods to estimate scatter distributions, along with convetional LBTE and MS-LBTE methods, with different numbers of spectra in CT scans in the same hardware environment (Nvidia RTX A5000, $N_{V}=16^3$, $N_{E}=10$ and $N_{A}=114$). 
	In terms of computation accuracy, we used GEANT4 to conduct MC simulations to generate primary and scatter signals at different spectra as references. The detector consisted of $80\times80$ pixels with a pixel size of $5.4$ mm; seven different spectra were used in MC simulations; the angular flux of X-ray source was set to $4.37142\times 10^9$ $\mathrm{sr^{-1}}$. 
	The source-to-isocenter distance (SID) and the source-to-detector distance (SDD) were $750.2$ $\mathrm{mm}$ and $1184.04$ $\mathrm{mm}$, respectively, with 7.4$^\circ$ cone angle of the X-ray source. In MSD-LBTE, $N_{E} = 9$, $N_{V} = 24^3$ and $N_{A}=114$. The phantoms used were a water cylinder with a diameter in $200$ $\mathrm{mm}$ and a height in $200$ $\mathrm{mm}$, and a voxelized head phantom consisting of water and bone with different densities.
	
	To further validate efficiency of the unified scatter correction method based on our proposed MSD-LBTE, we conducted a set of physical experiments, on an experimental CBCT tabletop system with a Varian 4030 DX flat-panel detector. 
	The Gammex$^{\mathbf{TM}}$ multi-energy CT phantom and the Kyoto head phantom were used in our experiments. The kernel-based method\cite{sunImprovedScatterCorrection2010}, and the CBCT Software Tools (CST) \cite{maslowskiAcurosCTSFast2018,wangAcurosCTSFast2018} with embedded LBTE-based scatter correction were also conducted as comparison. Some key parameters are listed in TABLE \ref{tab:setup}, with kV-switching mimicked by dual-kVp scans at 80/140 kVp for the Multi-Energy phantom and at 80/120 kVp for the head phantom, respectively.
	
	\begin{table}[htbp!]
		\centering
		\caption{Experiment Setup}
		\begin{tabular}{c|c|cc}
			\hline
			Experiment &kV-modulation &\multicolumn{2}{c}{kV-switching} \\
			\hline
			phantom &water Phantom &Multi-Energy Phantom &Head Phantom \\
			\hline 
			SID         &750mm          &750mm           &755mm  \\
			SDD         &1184mm         &1184mm          &1155mm \\
			Pixel size  &$0.45$mm       &$0.388$mm       &$0.388$mm \\
			Detector    &$960\times960$ &$1024\times768$ &$1024\times768$ \\
			Voltage     &70 to 140kVp   &80/140kVp       &80/120kVp \\
			Filter      &2mm Al         & 2mm Al         &0.3mm Cu \\
			\hline 
		\end{tabular}
		\label{tab:setup}
	\end{table}
	
	In addition, we also conducted a numerical simulation to explore potential of applying MSD-LBTE in kV-modulation enabled CT \cite{zouInvestigationCalibrationPhantoms2013,suMultimaterialDecompositionSpectral2022}. Here, a virtual kV-modulation CBCT scan was carried out, using an elliptical cylinder whose long axis is set as $24$ cm, short axis as $12$ cm, and height as $20$ cm. Projection data with and without scatter signals were generated with a kV-modulation designed to optimize the SNR over dose \cite{shinRadiationDoseReduction2013}. Some key parameters are also listed in TABLE \ref{tab:setup}, with X-ray source varying in the range of 70 to 140 Kilo-voltage peak (kVp) with an interval of 0.5 keV. Meanwhile, in order for comparison, we also approximate the continuously varying tube voltages to one fixed energy level of 105 kVp, and two fixed energy levels of 87.5 and 122.5 kVp, respectively, for the conventional LBTE-based scatter estimation and correction.
	
	\subsection{Evaluation metrics}
	
	In the comparison of scatter estimation accuracy, the mean relative error (MRE) and the structural similarity index (SSIM) are calculated in the images of scatter signals. The MRE is defined as,
	\begin{equation}
		\begin{aligned}
			MRE = \frac{1}{N_D}\sum_{i=1}^{N_D}|\frac{I^{ref}_i - I_i}{I^{ref}_i}|,
		\end{aligned}
	\end{equation}
	where $I^{ref}_i$ is the $i$-th pixel value of the reference image, and $I_i$ is the $i$-th pixel value of the obtained image. SSIM values were calculated by, 
	
	\begin{equation}
		\begin{aligned}
			SSIM = \frac{(2\mu_x\mu_y+C_1)(2\sigma_{xy}+C_2)}{(\mu_x^2+\mu_y^2+C_1)(\sigma_x^2+\sigma_y^2+C_2)}
		\end{aligned}
	\end{equation}
	where, $\mu_x$, $\mu_y$, $\sigma_x$, $\sigma_y$ and $\sigma_{xy}$ represent the local mean, standard deviation and cross-covariance of images x and y. In addition, in order to avoid errors due to the dynamic range difference,, we normalized the scatter signals to [0,1] when calculating SSIM.
	
	For the numerical simulations and physical experiments, we use an average error of water or iodine density among the regions of interest (ROIs) ($\overline{RMSE}$) defined as,
	\begin{equation}
		\begin{aligned}
			\overline{RMSE} = \frac{1}{M}\sum_{i=1}^{M}\sqrt{\frac{1}{N_{ROI_i}}\sum_{j=1}^{N_{ROI_i}}(I^{ref}_{i:j} - I_{i;j})^2} 
		\end{aligned}
	\end{equation}
	
	Where, $I^{ref}_{i;j}$ is the $j$-th pixel value of the $i$-th ROI of the reference image, $I_{i;j}$ is the $j$-th pixel value of the $i$-th ROI of the obtain image. Likewise, we use an average peak signal-to-noise ratio of water or iodine density in ROIs ($\overline{PSNR}$) defined as,
	\begin{equation}
		\begin{aligned}
			\overline{PSNR} = \frac{1}{M}\sum_{i=1}^{M}\left\{20\cdot \lg(\frac{\max_{j \in \rm{ROI_i}}(I_{i;j})}{RMSE_i})\right\}
		\end{aligned}
	\end{equation}
	where $RMSE_i$ is the root mean square error of the $i$-th ROI. 
	
	\section{Results}\label{sec: results}
	
	\subsection{Computational Efficiency}
\begin{figure} [htbp!]
		\begin{center}
			\begin{tabular}{c} 
				\includegraphics[height=0.5\linewidth]{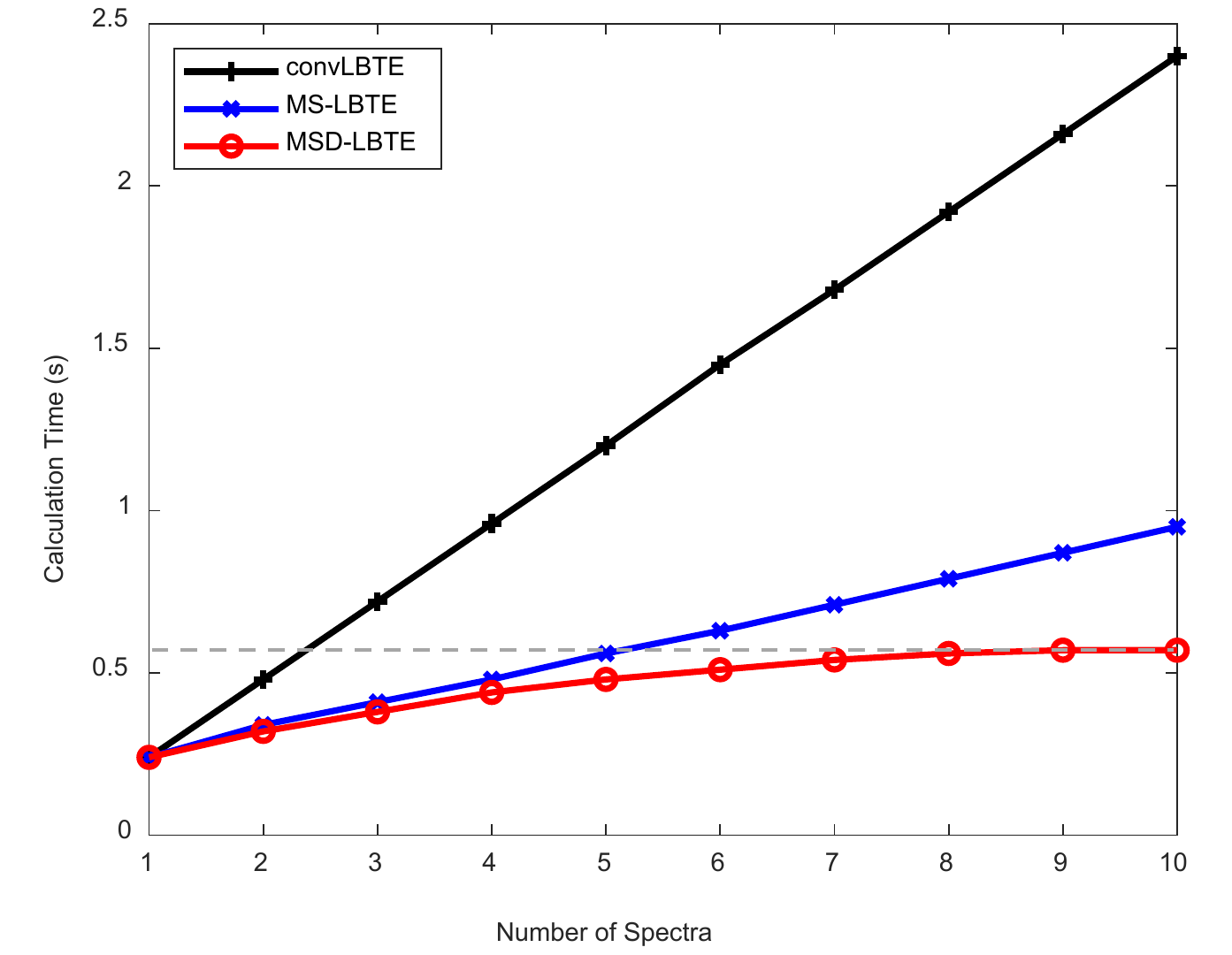}
			\end{tabular}
		\end{center}
		\caption[example] 
		{
			The computation time of scatter estimation by using different LBTE-based methods for different number of spectra when $N_{E}=9$.
		}
		\label{fig:9_TimeCost} 
	\end{figure}

	The actual computational costs for different number of spectra are shown in Fig. \ref{fig:9_TimeCost} with energy group $N_E = 9$. For the conventional LBTE-based scatter correction (convLBTE), the computation time increases linearly as the number of spectra increases, as expected, with about $0.24$ s per spectrum. For MS-LBTE, it still increases linearly but computation time per spectrum is decreased by a factor of two-third, thanks to the sharing of common calculations across different spectra being considered. 
	
	Using our proposed MSD-LBTE, the computational cost further reduced significantly even compared with MS-LBTE. It is observed that as the number of spectra increases, the computation time will reaches a saturation point of maximum limit. In the case of two spectra, the computation time can be reduced to 66.7\% of that of the conventional LBTE. The number drops to 52.8\% for three spectra. When 9 spectra used, the computation time of MSD-LBTE saturates, which is equivalent to 26.5\% of that of the conventional LBTE.

	\subsection{Computational Accuracy}

	\begin{table}[htbp!]
		\centering
		\caption{Accuracy of scatter distributions between Geant4 and MSD-LBTE for different spectra in Figs. \ref{fig:7_SpecCylin},\ref{fig:8_SpecHead}}
		\begin{tabular}{c|cc|cc}
			\hline
			\multirow{2}{*}{Spectrum} &\multicolumn{2}{c|}{Cylinder} &\multicolumn{2}{c}{Head} \\
			&MRE(\%) &SSIM &MRE(\%) &SSIM \\
			\hline 
			Spec1 &2.39 &0.9741 &2.94 &0.9658 \\
			Spec2 &1.13 &0.9847 &1.03 &0.9769 \\
			Spec3 &1.81 &0.9796 &1.79 &0.9723 \\
			Spec4 &1.32 &0.9825 &1.34 &0.9739 \\
			Spec5 &1.99 &0.9789 &2.40 &0.9718 \\
			Spec6 &2.38 &0.9668 &2.90 &0.9546 \\
			Spec7 &2.28 &0.9731 &2.33 &0.9617 \\
			\hline 
		\end{tabular}
		\label{tab:Mc_MRE}
	\end{table}
	
\begin{figure} [htbp!]
		\begin{center}
			\begin{tabular}{c}
				\includegraphics[height=0.7\linewidth]{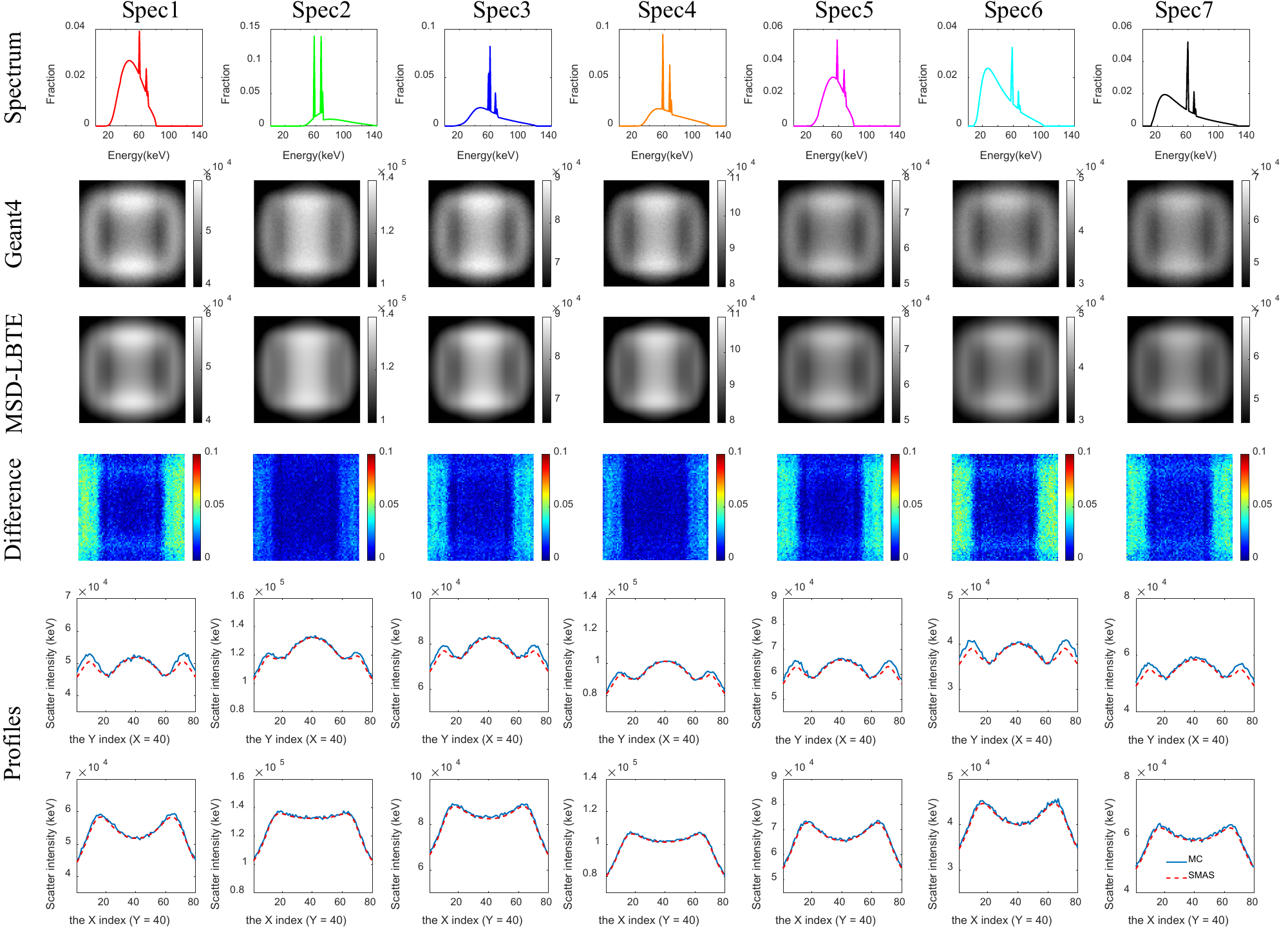}
			\end{tabular}
		\end{center}
		\caption[example] 
		{
			Scatter signals of water cylinder phantom for multiple spectra computed by Geant4 and our proposed MSD-LBTE. Spectrum: seven different spectra used; Geant4: the scatter signals estimated by Geant4; MSD-LBTE: the scatter signals estimated by MSD-LBTE; Difference: the relative errors between the scatter signals estimated by the two methods; and Profile: the central horizontal and vertical profiles of the scatter signals.
		}
		\label{fig:7_SpecCylin} 
	\end{figure}

\begin{figure} [htbp!]
		\begin{center}
			\begin{tabular}{c}
				\includegraphics[height=0.7\linewidth]{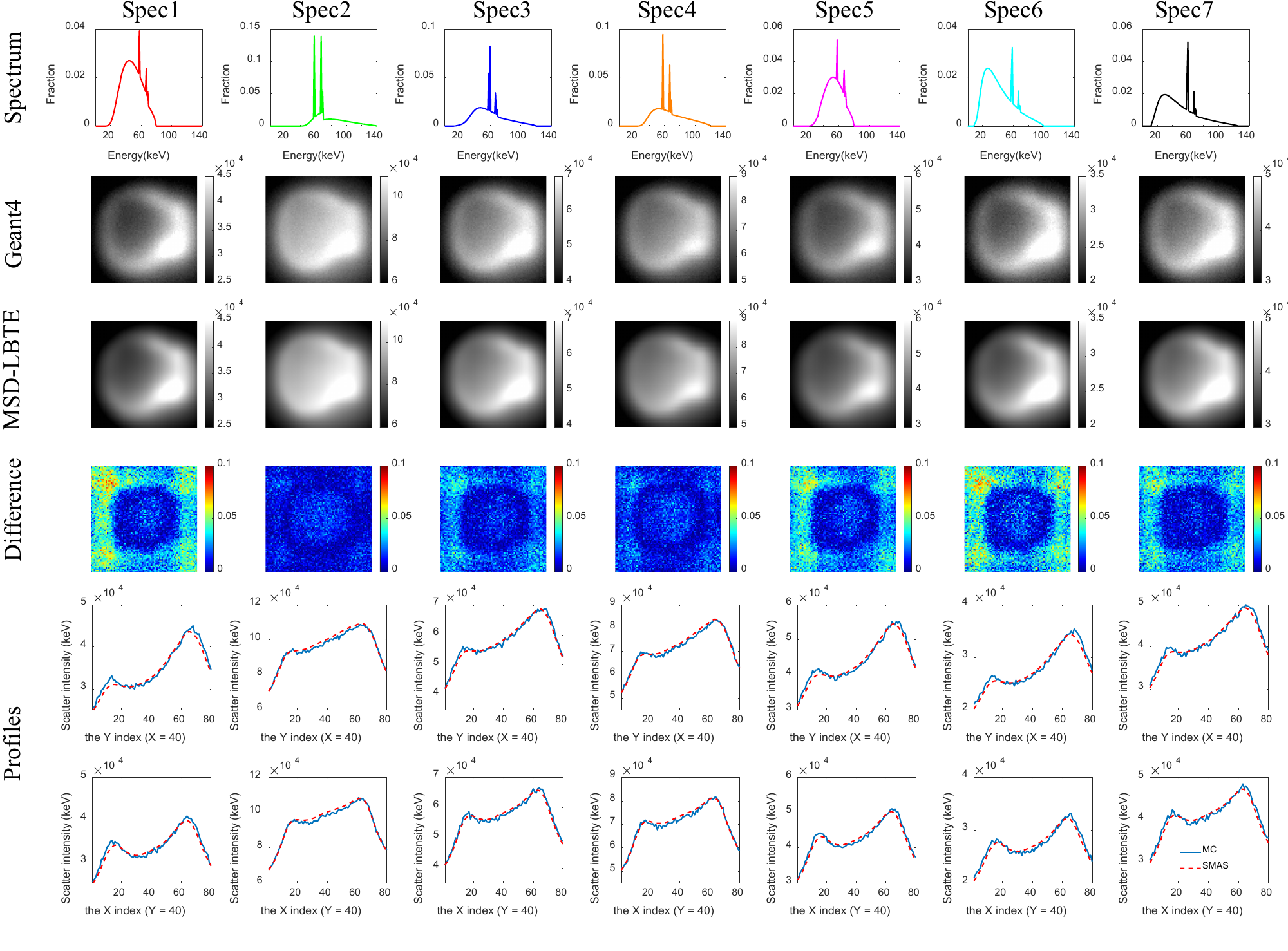}
			\end{tabular}
		\end{center}
		\caption[example] 
		{
			Scatter signals of voxelized head phantom for multiple spectra computed by Geant4 and our proposed MSD-LBTE. Spectrum: seven different spectra used; Geant4: the scatter signals estimated by Geant4; MSD-LBTE: the scatter signals estimated by MSD-LBTE; Difference: the relative errors between the scatter signals estimated by the two methods; and Profile: the central horizontal and vertical profiles of the scatter signals.
		}
		\label{fig:8_SpecHead} 
	\end{figure}

	In Figs. \ref{fig:7_SpecCylin} and \ref{fig:8_SpecHead}, we compare the simulated scatter signals of the MC simulation with that of MSD-LBTE, using the water cylinder phantom and the voxelized head phantom, respectively, where the horizontal and vertical line profiles across the center of the primary and scatter signal images are also plotted. It is seen that overall results of MSD-LBTE agrees well with the MC simulation. The MRE and SSIM of the scatter signals are summarized in TABLE \ref{tab:Mc_MRE}. Quantitatively, for the water cylinder phantom, MRE is less than $2.39\%$ and SSIM is greater than $96.68\%$ for every spectrum, respectively; while for the voxelized head phantom, they are less than $2.94\%$ and greater than $95.46\%$ across all spectra, respectively.
	
	It is worth noting that although our proposed MSD-LBTE achieved a comparable accuracy to the MC method, in scatter computation, with much faster speed, some noticeable discrepancies between MC and MSD-LBTE appear mainly density in the peripheries. This is because that the sampling (discretization) is not dense enough in those regions, where the primary/scatter signal changes sharply. Increasing the discretization density (such as from $N = 24^3$ to $N = 32^3$) can reduce those discrepancies. Fortunately, however, they will not affect scatter correction much in the end, as the primary signals are relatively much bigger (resulting in a negligible residual error in SPR). 
	
	\subsection{Scatter Correction in KV-Switching Spectral CT scan}
	
	\subsubsection{Multi-Energy phantom study}

	\begin{figure}[htbp!]
		\centering
		\includegraphics[width=0.7\linewidth]{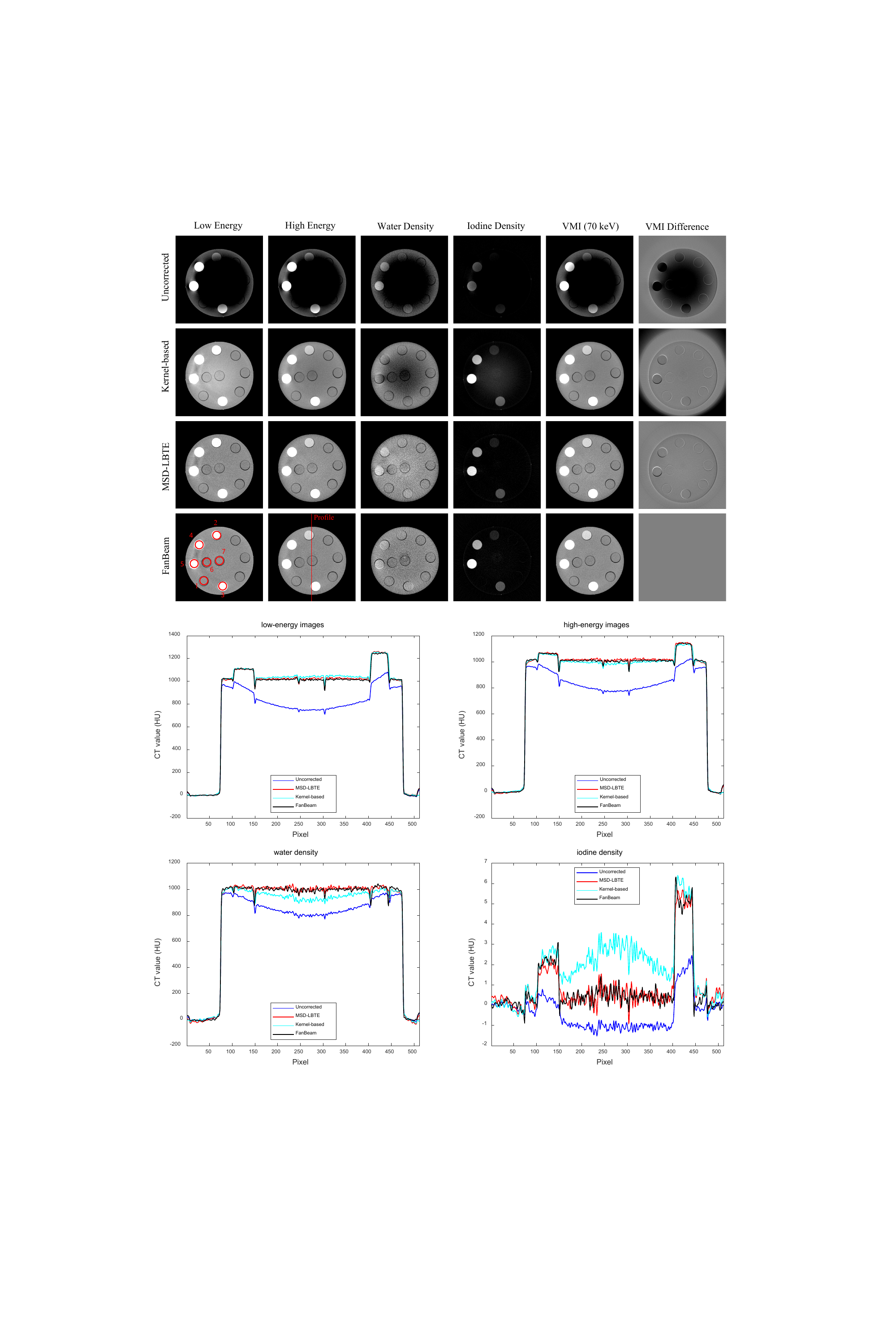}
		\caption{Reconstructions of the multi-energy CT phantom: low- and high-energy reconstruction images, material decomposition images, and their profiles. The display windows is [900 1100] HU for low- and high-energy reconstruction images and virtual monoenergetic image (VMI) at $70$ keV, [0.9 1.1]$\mathrm{g/cm^3}$ for water density, and [0 15]$\mathrm{mg/ml}$ for iodine density.}
		\label{fig:10_Gam}
	\end{figure}
	
	The reconstructed CT images from low- and high-energy projections, as well as basis material projections after material decomposition after material decomposition are shown in Fig. \ref{fig:10_Gam}.
	It is seen that, with scatter correction there are serious artifacts in the CT images  as expected, and the water and iodine densities are way off their nominal values. Our proposed MSD-LBTE method significantly removes the artifacts and recoveries water and iodine densities. The plotted profiles show that after scatter correction, the CT values from cone-beam scans, along with their water and iodine densities after material decomposition are in good agreement with the corresponding fan-beam scan (reference).
	
	\begin{table}[htbp!]
		\centering
		\setlength{\tabcolsep}{2pt}
		\caption{ Averaged density (mg/ml), $\overline{RMSE}$ and $\overline{PSNR}$ for ROIs of water and iodine density images in Fig. \ref{fig:10_Gam}.}
		\label{tab:md_gammax_table}
		\begin{tabular}{c|ccccc}
			\hline
			\multirow{2}{*}{ROI} &\multicolumn{5}{c}{Iodine Density (mg/ml)} \\
			\cline{2-6}
			&Ground truth &Uncorrected &Kernel-based &MSD-LBTE &Fan beam \\
			\hline 
			1          &0.0  &-0.3 &0.7   &{0.1}   &0.1   \\
			2          &2.0  &0.4  &2.5   &{2.0}   &2.0   \\
			3          &5.0  &1.7  &5.7   &{5.2}   &5.1   \\
			4          &10.0 &3.8  &10.6  &{9.7}   &10.0  \\
			5          &15.0 &5.6  &15.7  &{14.9}  &14.9  \\
			6          &0.0  &-1.1 &2.9   &{0.5}   &0.6   \\
			7          &0.0  &-1.2 &2.4   &{0.6}   &0.4   \\
			$\overline{RMSE}$ &0.0  &3.30 &1.33  &\textbf{0.47}  &0.36  \\
			$\overline{PSNR}$ &-    &0.98 &13.73 &\textbf{18.76} &20.20 \\
			\hline 
			\multirow{2}{*}{ROI} &\multicolumn{5}{c}{Water Density (mg/ml)} \\
			\cline{2-6}
			&Ground truth &Uncorrected &Kernel-based &MSD-LBTE &Fan beam \\
			\hline 
			1          &1000.0 &919.8  &994.9 &{1012.1} &1010.7 \\
			2          &1000.0 &936.3  &997.4 &{1018.3} &1017.0 \\
			3          &1000.0 &953.4  &995.6 &{1019.0} &1023.0 \\
			4          &1000.0 &986.4  &997.5 &{1036.6} &1031.0 \\
			5          &1000.0 &1019.6 &993.1 &{1037.3} &1036.7 \\
			6          &1000.0 &802.8  &921.7 &{1004.7} &991.4  \\
			7          &1000.0 &834.8  &943.4 &{999.6}  &995.7  \\
			$\overline{RMSE}$ &0.0    &87.67  &31.07 &\textbf{23.09}  &21.65  \\
			$\overline{PSNR}$ &-      &23.08  &32.71 &\textbf{34.03}  &34.60  \\
			\hline 
		\end{tabular}
	\end{table}
	
	Quantitatively, TABLE \ref{tab:md_gammax_table} lists the averaged density, $\overline{RMSE}$ and $\overline{PSNR}$ in the selected ROIs, where the groundtruth of iodine and water density are referred to the user manual of the Gammex Multi-Energy CT phantom.
	It is seen that the results from uncorrected cone-beam scan have significant bias. After scatter correction by our proposed method, in iodine density, the $\overline{RMSE}$ is reduced from $3.30$ mg/ml to $0.47$ mg/ml and the $\overline{PSNR}$ is increased from $0.98$ dB to $18.76$ dB; in water density, the $\overline{RMSE}$ is reduced from $87.67$ mg/ml to $23.09$ mg/ml and the $\overline{PSNR}$ is increased from $23.08$ dB to $34.03$ dB. 
	Besides, compared with the conventional LBTE method of separately conducting scatter correction for high and low energy in dual-energy kV-switching CT, MSD-LBTE reduces the computation time by $1/3$.

	\subsubsection{Head phantom study}
	
	\begin{figure}[htbp!]
		\centering
		\includegraphics[width=0.7\linewidth]{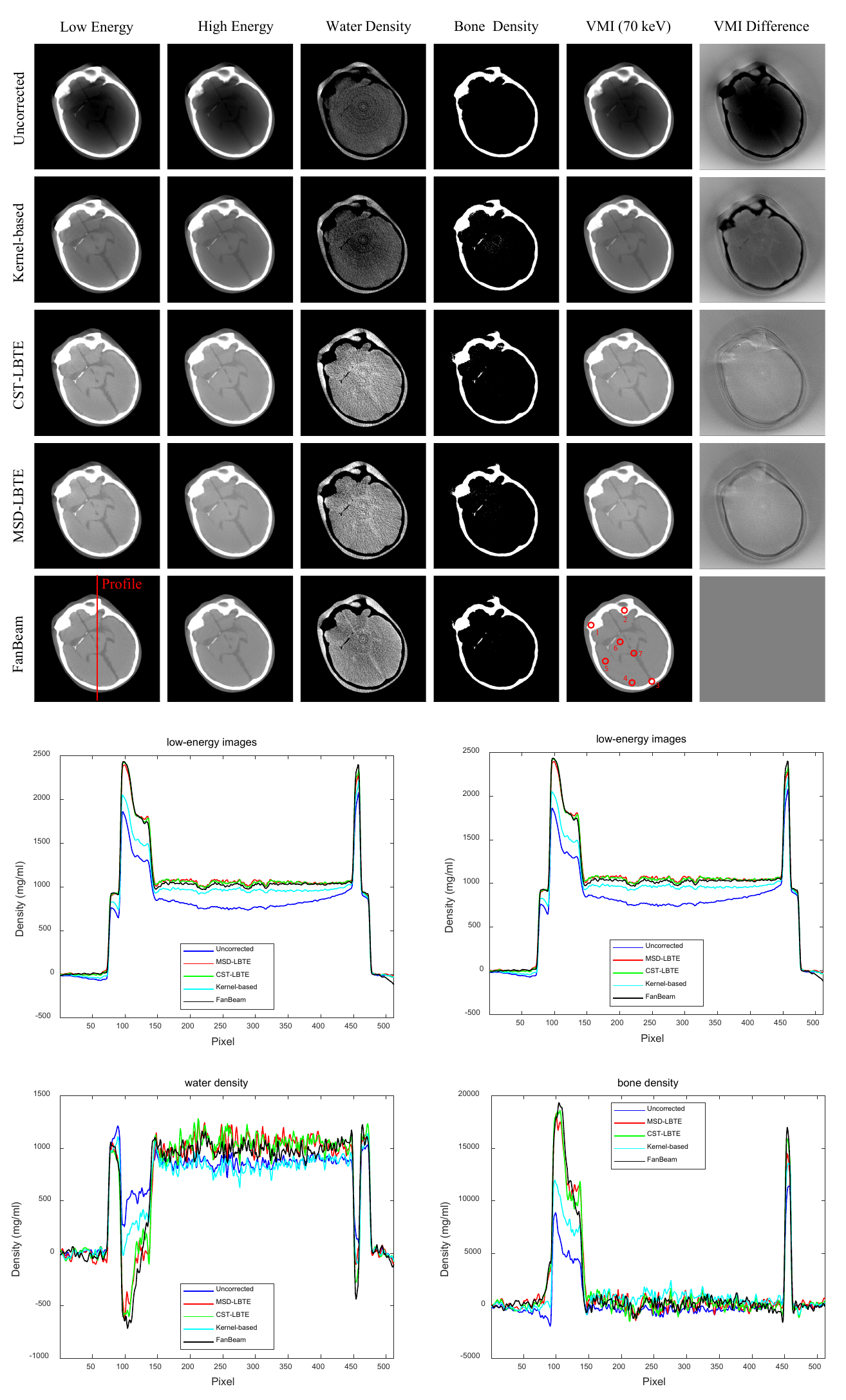}
		\caption{Reconstructions of head CT phantom: low- and high-energy reconstruction images, material decomposition Images, and their profiles. The display windows is [750 1250] HU for low- and high-energy reconstruction images and virtual monoenergetic image (VMI) at $70$ keV, [-200 200] HU for VMI difference, [0.75 1.25]$\mathrm{g/cm^3}$ for water density, and [1.5 2.5]$\mathrm{g/cm^3}$ for bone density.}
		\label{fig:md_head}
	\end{figure}

	\begin{table}[htbp!]
		\centering
		\setlength{\tabcolsep}{2pt}
		\caption{Averaged CT value (HU), $E_{RMES}$ adn $\overline{PSNR}$ for ROIs of VMIs (70 keV) in Fig. \ref{fig:md_head}.}
		\begin{tabular}{c|ccccc}
			\hline
			\multirow{2}{*}{ROI} &\multicolumn{5}{c}{VMI (HU)} \\
			\cline{2-6}
			&Uncorrected &Kernel-based &CST-LBTE &MSD-LBTE &Fan beam \\
			\hline
			1          &1566.6 &1706.8 &2002.8 &{1975.0} &1957.0 \\
			2          &1252.1 &1435.6 &1765.1 &{1746.5} &1699.2 \\
			3          &1615.3 &1731.6 &1903.0 &{1871.0} &1925.8 \\
			4          &917.0  &975.6  &1038.4 &{1034.1} &1024.9 \\
			5          &863.9  &952.2  &1048.0 &{1048.2} &1024.1 \\
			6          &783.2  &952.1  &1062.8 &{1073.8} &1025.3 \\
			7          &772.3  &915.0  &1022.5 &{1026.3} &987.1  \\
			$\overline{RMSE}$ &268.72 &140.61 &39.72  &\textbf{37.07}  &0      \\
			$\overline{PSNR}$ &13.15  &20.72  &31.80  &\textbf{32.53}  &-      \\
			\hline
		\end{tabular}
		\label{tab:vmi_head}
	\end{table}
	
	The low- and high-energy reconstruction image along with material decomposition results using different scatter correction methods and their profiles are shown in Fig. \ref{fig:md_head}. 
	It also shows the difference images between cone-beam VMIs with or without scatter correction with respect to the fan-beam VMI. 
	Preliminary results show that the CT reconstructions after scatter correction by using either CST or our proposed method are in good agreement with that from the fan-beam scans.
	Quantitatively, TABLE \ref{tab:md_gammax_table} lists the averaged CT values, $\overline{RMSE}$ ,and $\overline{PSNR}$ of the selected ROIs of VMIs, where the fan-beam scan results are utilized as a reference.
	Both CST and our proposed method show a good performance in scatter correction, demonstrating the effectiveness and accuracy of the LBTE-type method.

	\subsection{Potential in KV-Modulation CT scan}
	
	\begin{figure} [htbp!]
		\begin{center}
			\begin{tabular}{c}      
				\includegraphics[height=0.5\linewidth]{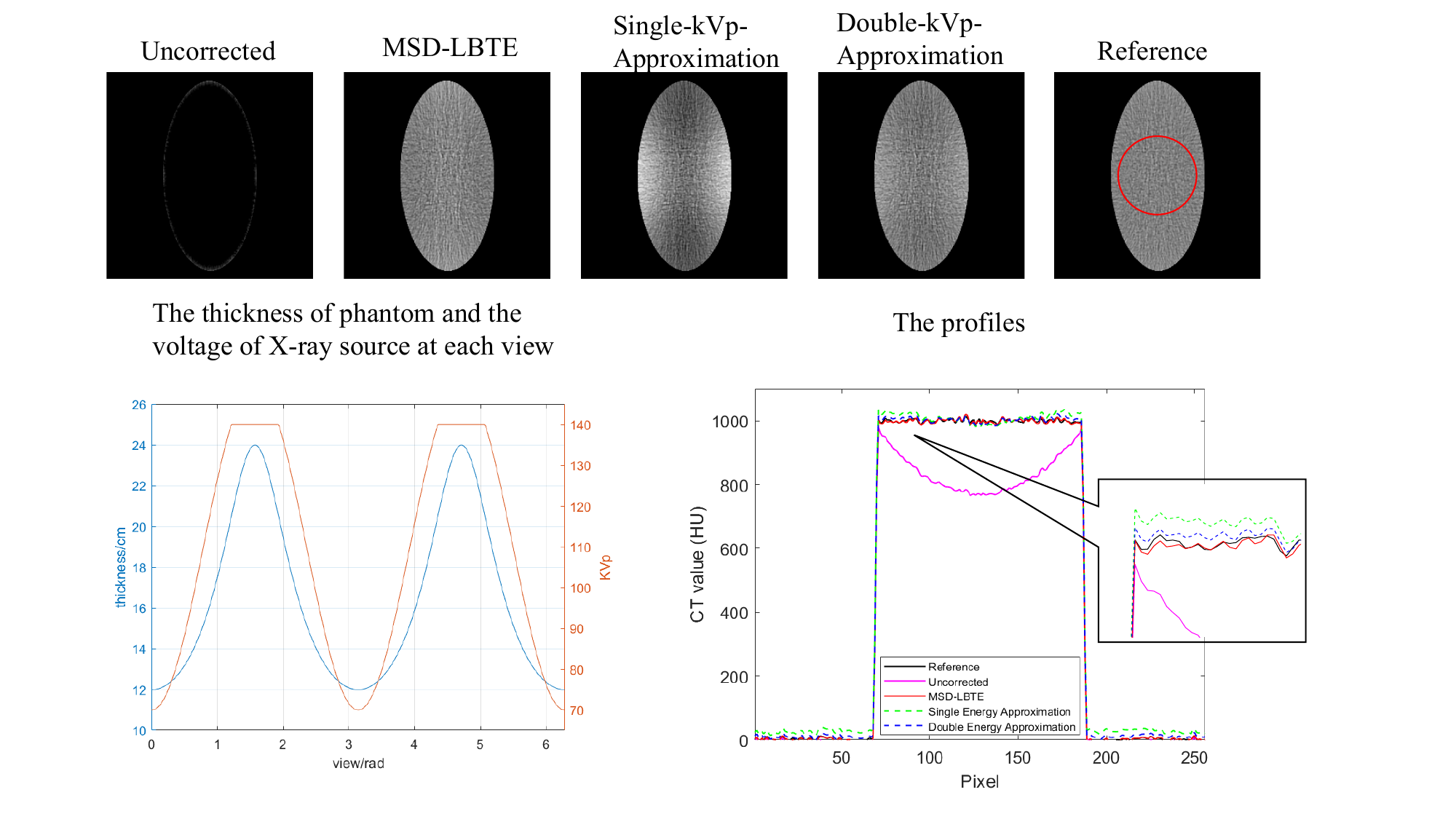}
			\end{tabular}
		\end{center}
		\caption[example] 
		{kV-modulation study: reconstruction image results with or without scatter correction (The red area are selected ROI) and their profiles. The display windows is [950 1050] HU.}
		\label{fig:9_5_kvModulation} 
	\end{figure} 
	
	As shown in Fig. \ref{fig:9_5_kvModulation}, before scatter correction, there are serious scatter artifacts in CT images reconstructed from the kV-modulated projections with $RMSE=187.6$, significantly deviating from the reference, which emphasizes the need of a proper scatter correction. After single-kVp-approximated LBTE correction which costs 0.24s (each view), the $RMSE$ is reduced to $11.41$ but residual scatter artifacts are quite obvious. Double-kVp-approximated LBTE correction which costs 0.48s further improves the image quality, with the $RMSE$ decreased to $5.89$ while some residual scatter artifacts still noticeable. 
	Using our proposed MSD-LBTE which costs 0.57s, the reconstructed CT images have the best uniformity visually and the lowest $RMSE=4.78$, with no obvious scatter artifacts. Compared with double-kVp-approximation LBTE correction, the computational time of using MSD-LBTE, which obtains scatter signals for all kVps without approximation, just increases 18\% overall.
	It is seen that, with our proposed method, scatter correction will no longer be a bottle-neck for CT scan with kV-modulation, in terms of computation at least.

	\section{Discussions and Conclusion}\label{sec: discussions}
	
	In this work, we proposed a matrixed-spectrum decomposition accelerated linear Boltzmann transport equation solver, and established a unified and enhanced scatter correction basing on it. Our proposed method significantly improves computational efficiency at multiple spectra while preserving high levels of accuracy.
	
	For our implementation of the LBTE method, while the computational load introduced by the phase space expansion is substantially small portion of the entire runtime, achieving the desired optimization is hindered by the time-consuming IO read and write conflicts in multi-threaded computer programming. Further balancing the IO read and write operations to mitigate access conflicts in multi-thread process could further improve the acceleration in computation speed.
	Moreover, the MSD-LBTE can be applied not only in multi-spectral CT but also in multi-source CT configurations such as stationary CT, where $l$ can be employed to distinguish between different positions of sources with different spectra potentially.
	Also, our work could play a significant role for learning-based tasks where efficiently generating CT scans from publicly available CT datasets for a range of different kVp's, with realistic scatter signals, becomes more and more important.
	
	There exists some limitations for the LBTE-type method. 
	First, a challenge lies in performing scatter correction on truncated projection data, regardless of whether the truncation occurs in the $\mathrm{x-y}$ direction or the $\mathrm{z}$ direction. Portions of the scanned objects that are outside the feild of view (FOV) still contribute to the scatter signals. 
	Second, it is difficult to conduct scatter correction for objects with intricate structure composed of various materials with significant differences in attenuation and scattering characteristics. In such a case, the accuracy of material conversion could be notably compromised.
	These two issues exists for most post-reconstruction methods.
	
	Future endeavors includes further optimization of the implementation to enhance both computation speed and accuracy, exploration of new application, and more in-depth assessments and analyses in a variety of multi-spectral CT scenarios beyond fast-kV switching based spectral CT and kV-modulation enabled CT scans.
	
	\section{Acknowledgment}
	This project was supported in part by Grants from the National Key R\&D Program of China (No. 2022YFE0131100 and 2024YFC2417500), and in part by Beijing Natural Science Foundation (No. L252021).
	
	\bibliography{MSDBS}
	
\end{document}